\documentclass[acmsmall,screen,authorversion,nonacm]{acmart}

\AtBeginDocument{%
  }

\setcopyright{none}
\copyrightyear{2023}
\acmYear{2023}
\acmDOI{}

\acmConference[PLDI'23]{44th ACM SIGPLAN Conference on Programming Language Design and Implementation}{June 19--21, 2023}{Orlando, FL}

\startPage{1}

\setcitestyle{nocompress}   

\usepackage{booktabs}
\usepackage{subcaption}

\hypersetup{colorlinks,allcolors=blue}

\usepackage[boxed,vlined,ruled]{algorithm2e}
\SetAlCapSkip{0.5em}

\usepackage{amsmath}
\usepackage{stmaryrd}
\usepackage{enumitem}
\usepackage{wrapfig}
\usepackage{tikz}
\usetikzlibrary{fit}
\usepgflibrary{shapes.symbols}

\newcommand{\df}[1]{{\it #1}}
\newcommand{\Oh}{{\ensuremath{\mathcal{O}}}}

\newcommand{\NP}[0]{\texttt{NP}\xspace}
\newcommand{\alg}[1]{{\bf \texttt{#1}}\xspace}
\newcommand{\prob}[1]{\textsc{#1}\xspace}

\newcommand{\bps}{\alg{bps}}  
\newcommand{\bpc}{\alg{bpc}}  

\begin{document}

\title{Optimizing Function Layout for Mobile Applications}


\author{Ellis Hoag}
\email{ellishoag@meta.com}          
\orcid{0000-0003-3853-1889}             
\affiliation{
  \institution{Meta Platforms Inc.}            
  \country{USA}                    
}

\author{Kyungwoo Lee}
\orcid{0000-0002-9127-7261}             
\email{kyulee@meta.com}          
\affiliation{
  \institution{Meta Platforms Inc.}            
  \country{USA}                    
}

\author{Juli\'{a}n Mestre}
\orcid{0000-0003-4948-2998}             
\email{julianmestre@meta.com}         
\affiliation{
  \institution{Meta Platforms Inc.}           
  \country{USA}                   
}
\email{julian.mestre@sydney.edu.au}         
\affiliation{
  \institution{University of Sydney}           
  \country{Australia}                   
}

\author{Sergey Pupyrev}
\orcid{0000-0003-4089-673X}             
\email{spupyrev@meta.com}          
\affiliation{
  \institution{Meta Platforms Inc.}            
  \country{USA}                    
}



\begin{abstract}
  Function layout, also referred to as function reordering or function placement, is one of the 
  most effective profile-guided compiler optimizations. By reordering functions in a binary,
  compilers are able to greatly improve the performance of large-scale applications or  
  reduce the compressed size of mobile applications. Although the technique has been studied in the
  context of large-scale binaries, no recent study has investigated the impact of function layout on
  mobile applications.
  
  In this paper we develop the first principled solution for optimizing function layouts in the mobile
  space. To this end, we identify two important optimization goals, the compressed code size
  and the cold start-up time of a mobile application. Then we propose a formal model for the layout
  problem, whose objective closely matches the goals. Our novel algorithm to optimize the layout
  is inspired by the classic balanced graph partitioning problem. We carefully engineer and implement
  the algorithm in an open source compiler, LLVM. An extensive evaluation of the new method on
  large commercial mobile applications indicates up to $2\%$ compressed size reduction and up to
  $3\%$ start-up time improvement on top of the state-of-the-art approach.
\end{abstract}

\maketitle

\section{Introduction}
\label{sect:intro}

As mobile applications become an essential part of everyday life, the task of making them faster, smaller, and
more reliable becomes urgently important. Profile-guided optimization (PGO) is a critical component in modern compilers for improving the performance and the size of applications, which in turn enables the development and delivery of new
app features for mobile devices with limited storage and low memory. The technique, also known as 
feedback-driven optimization (FDO), leverages the program's dynamic behavior to generate optimized 
applications. Currently PGO is a part of most commercial and open-source compilers 
for static languages; it is also used for dynamic languages, as a part of Just-In-Time (JIT) compilation.

Modern PGO has been successful in speeding up server workloads~\cite{CML16,PANO19,HMPWY22} by  
providing a double-digit percentage performance boost. This is achieved via a combination of
multiple compiler optimizations which include function inlining, loop optimizations, and code layout.
PGO relies on program execution profiles, such as the 
execution frequencies of basic blocks and function invocations, to guide compilers to optimize 
critical paths of a program more selectively and effectively. Typically, server-side PGO targets at
improving CPU and cache utilization during the steady state of program execution, leading to higher server throughput.
This poses a unique challenge for applying PGO for mobile applications which are largely I/O bound and do not have a well-defined steady-state performance due to their user-interactive nature~\cite{LHT22}.
Instead, app download speed and app launch time are critical to the success of mobile apps because they directly 
impact user experience, and therefore, user retention~\cite{CLB21,LFZLS22}.

In this paper we revisit a classic PGO technique, function layout, and show how to successfully apply it
in the context of mobile applications. We emphasize that most of the earlier compiler optimizations 
focus on a single objective, such as the performance or the size of a binary. However, function layout
might impact multiple key metrics of a mobile application. We show how to place functions in a binary
to simultaneously improve its (compressed) size and start-up performance. The former objective
is directly related to the app download speed and has been extensively discussed in recent works on
compiler optimizations for mobile applications~\cite{LFZLS22,LRN22,CLB21,LHT22,RPFB22,RPWCL20}. The latter receives considerably
less attention but nevertheless is of prime importance in the mobile space~\cite{A22,G22,M20,F15}.

Function layout, along with basic block reordering and inlining, is among the most impactful 
profile-guided compiler optimizations. The seminal work of Pettis and Hansen~\cite{PH90} introduces
a heuristic for function placement that leads to a reduction in I-TLB (translation lookaside buffer) 
cache misses, which improves the steady-state performance of large-scale binaries. The follow up
work of Ottoni and Maher~\cite{OM17} describes an improvement to the placement scheme by
considering the performance of the processor instruction cache (I-cache). The two heuristics 
are implemented in the majority of modern compilers and binary
optimizers~\cite{PANO19,LCD19,OM17,Prop21}. However, these optimizations are not widely used
in mobile development and the corresponding layout algorithms have not been thoroughly studied.
To the best of our knowledge, the recent work of Lee, Hoag, and Tillmann~\cite{LHT22} is the 
only study describing a technique for function placement for native mobile applications.
With this in mind, we provide \emph{the first comprehensive investigation of function layout algorithms}
in the context of mobile applications. Next in Section~\ref{sect:bpc} we explain how 
function layout impacts the compressed app size, which eventually affects download speed. Then in 
Section~\ref{sect:bps} we describe how an optimized function placement can improve the start-up time.
Finally, Section~\ref{sect:cont} highlights our main contributions, \emph{a unified optimization model} 
to tackle these two seemingly unrelated objectives and \emph{a novel algorithm} for the problem
based on the recursive balanced graph partitioning.

\subsection{Function Layout for App Download Speed}
\label{sect:bpc}

As mobile apps continue to grow rapidly, reducing the binary size is crucial for application developers~\cite{LHT22,B21,LFZLS22}.
Smaller apps can be downloaded faster, which directly impacts user 
experience~\cite{R16}. For example, a recent study in \cite{CLB21} establishes a strong correlation between the app size and the user engagement.
Furthermore, mobile app distribution 
platforms may impose size limitations for downloads that use cellular data~\cite{CLB21}. Indeed, in the Apple App Store, when an app's size crosses a certain threshold, users will not get timely updates, which often include critical security improvements, until they are connected to a Wi-Fi network.

\begin{figure}[!tb]
  \centering
  \includegraphics[width=0.9\columnwidth,page=2]{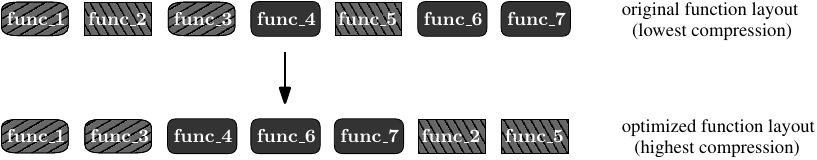}
  \caption{Placing similar (same-patterned) functions nearby in the binary leads to higher 
    compression rates achieved by Lempel-Ziv algorithms. Functions are considered similar when
    they share common sequences of instructions that can be encoded by short references.}
  \label{fig:compression}
\end{figure}

Mobile apps are distributed to users in a compressed form via the mobile app platforms. Typically, application
developers do not have a direct control over the compression technique utilized by the platforms. However, the
recent work of Lee, Hoag, and Tillmann~\cite{LHT22} observes that by modifying the content of a binary, one
may improve its compressed size. In particular, co-locating ``similar'' functions in the binary improves the
compression ratio achieved by popular compression algorithms such as \textsf{ZSTD} or \textsf{LZFSE}.
A similar technique, co-locating functions based on their similarity, is utilized in Redex, a bytecode
Android optimizer~\cite{Redex}.

Why does function layout affect compression ratios? Most modern lossless compression tools rely on the
Lempel-Ziv (LZ) scheme~\cite{LZ77}. Such algorithms try to identify long repeated sequences in the data and
substitute them with pointers to their previous occurrences. If the pointer is represented in fewer bits than 
the actual data, then the substitution results in a compressed-size win. That is, the shorter the distance
between the repeated sequences, the higher the compression ratio is.
To make the computation effective, LZ-based algorithms search for common sequences inside a \df{sliding window},
which is typically much shorter than the actual data. Therefore, function layouts in which repeated
instructions are grouped together, lead to smaller (compressed) mobile apps; 
see Figure~\ref{fig:compression} for an example.
In Section~\ref{sect:model} we describe an algorithmic framework for finding such layouts. 

\subsection{Function Layout for App Launch Time}
\label{sect:bps}

Start-up time is one of the key metrics for mobile applications. Launching an app
quickly is important to ensure users have a good first impression. Start-up
delays have a direct impact on user engagement~\cite{YCG12}. A study in \cite{M20}
indicates that $20\%$ of users abandon an app after one use and $80\%$ of them give a poorly performing app 
three chances or fewer before uninstalling it.

Start-up time is the time between a user clicking on an application icon to the display of the first
frame after rendering. There are several start-up scenarios: cold start, warm start, and hot start~\cite{A22,G22,F15}.
Switching back and forth between different apps on a mobile leads to hot/warm start and typically
does not incur significant delays. In contrast, starting an app from scratch or resuming it after a memory
intensive process is referred to as cold start. Our main focus is on improving this cold start scenario,
which is usually the key performance metric.

\begin{figure}[!tb]
  \centering
  \includegraphics[width=0.9\columnwidth,page=1]{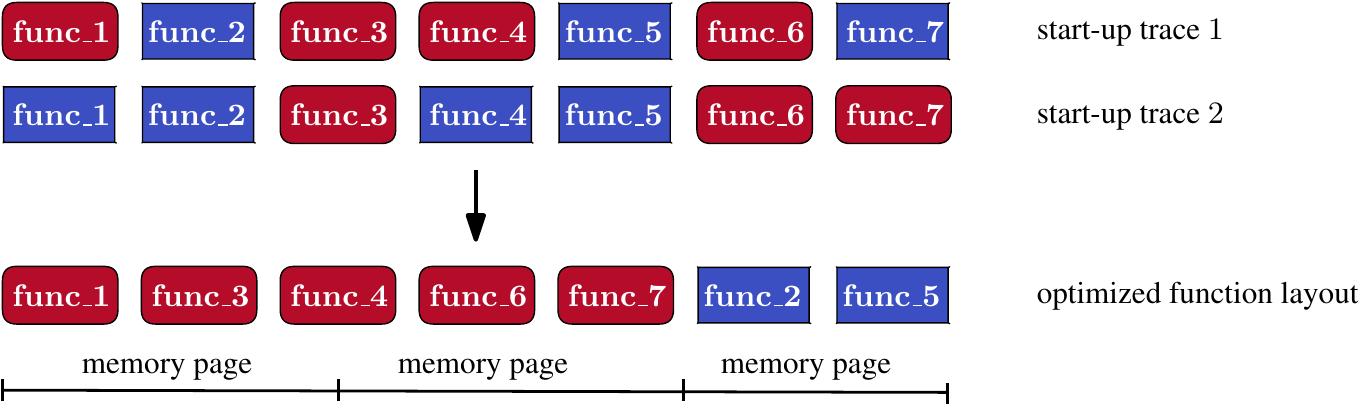}
  \caption{Co-locating hot (round) functions and cold (rectangular) functions nearby in the binary
    leads to a reduction in page faults. The hotness of the functions and their order of executions
  might depend on the usage scenario (trace), and the task is to find a single optimized function layout.}
  \label{fig:faults}
\end{figure}

Unlike server workloads, where code layout algorithms optimize the cache utilization~\cite{OM17,PANO19,NP20}, 
start-up performance is mostly dictated by memory page faults~\cite{EF78}. When an app is launched, 
its code has to be transferred from disk to the main memory before it can be executed.
Function layout can affect performance because the transfer happens at the granularity of memory pages.
As illustrated in Figure~\ref{fig:faults}, interleaving cold functions that are never executed with hot ones
results in more memory pages to fetch from disk. A tempting solution is to group together hot functions in the
binary. Notice however that some mobile apps have a billion of daily active users who use the apps on a variety of devices and 
platforms. Thus, optimizing the layout for one usage scenario might result in sub-optimal performance for others.
The main challenge is to produce a single function layout that optimizes the start-up performance
across all use cases.
This is accomplished by a novel optimization model, which we develop in Section~\ref{sect:model}.

\subsection{Contributions}
\label{sect:cont}

We model the problem of computing an optimized function layout for mobile apps as the 
balanced graph partitioning problem~\cite{GJ74}. The approach allows for a single algorithm that can improve both 
app start-up time (impacting user experience) and app size (impacting download speed).
However, while the layout algorithm is the same for the objectives, it operates with different data collected
while profiling the binary. For the sake of clarity, we call the optimizations 
\prob{Balanced Partitioning for Start-up Optimization}~(\bps) and \prob{Balanced Partitioning for Compression Optimization}~(\bpc).
Though we emphasize that the underlying implementation, outlined in Algorithm~\ref{algo:bp}, is identical.

The former optimization, \bps, is applied for \df{hot} functions in the binary that are executed during app start-up.
The latter optimization, \bpc, is applied for all the remaining \df{cold} functions. In our experiments, we 
record around $15\%$ of functions being hot. That allows us to reorder most of the functions in a ``compression-friendly''
manner, while at the same time improving the overall start-up performance. Compared to the prior work~\cite{LHT22}, we improve the compressed size by up to $2\%$ and the start-up time by up to $5\%$,
while speeding up the function layout phase by $20$x for 
\texttt{SocialApp}, one of the largest mobile apps in the world.
We summarize the contributions of the paper as follows.

\begin{itemize}
  \item We formally define the function layout problem in the context of mobile applications. To this end,
  we identify and formalize two optimization objectives, based on the application start-up time and
  the compressed size.
  
  \item Next we present the \prob{Balanced Partitioning} algorithm, which takes as input a bipartite graph between \df{function} and \df{utility} vertices, and outputs an order of the function vertices.
  We also show how to reduce the aforementioned objectives of \bpc and \bps to an instance for the balanced
  partitioning.
  
  \item Finally, we extensively evaluate the compressed size, the start-up performance, and the build time
  of the new algorithms with two large commercial iOS applications, \texttt{SocialApp} and \texttt{ChatApp}.
  Furthermore, we experiment with app size on Android native binaries, \texttt{AndroidNative}.
\end{itemize}

The rest of the paper is organized as follows.
Section~\ref{sect:model} builds an optimization model for compression and start-up performance, respectively.
Then Section~\ref{sect:alg} introduces the recursive balanced graph partitioning algorithm that is the basis to 
effectively solve the two optimization problems.
Next, in Section~\ref{sect:impl}, we describe our implementation of the technique in an open source compiler, LLVM.
Section~\ref{sect:eval} shows an experimental evaluation with real-world mobile applications.
We conclude the paper with a discussion of related works in Section~\ref{sect:related} and 
propose possible future directions in Section~\ref{sect:conclude}.

\section{Building an Optimization Model}
\label{sect:model}

We model the function layout problem with a bipartite graph, denoted $G=(F \cup U, E)$, where
$F$ and $U$ are disjoint sets of vertices and $E$ are the edges between the sets. The set $F$ is a collection
of all \df{functions} in a binary, and the goal is to find a permutation (also called an \df{order} or a \df{layout})
of $F$. The set $U$ represents auxiliary \df{utility} vertices that are used to define an objective for optimization.
Every utility vertex $u \in U$ is adjacent with a subset of functions, $f_1, \dots, f_k \in F$ so that
$(u, f_1), \dots, (u, f_k) \in E$ for some integer $k \ge 2$. Intuitively, the goal of the layout algorithm
is to place all functions so that $f_1, \dots, f_k$ are nearby in the resulting order, for each utility vertex $u$.
That is, the utility vertex encodes a locality preference for the corresponding functions.
Next we formalize the intuition for each of the two objectives.

\subsection{Compression}
\label{sect:model:compression}
As explained in Section~\ref{sect:bpc}, the compression ratio of a Lempel-Ziv-based algorithm can be improved if 
similar functions are placed nearby in the binary. This observation is based on earlier theoretical studies~\cite{RRRS13} 
and has been verified empirically~\cite{FM10,LLDSW14} in the context of lossless data compression. The works define 
(sometimes, implicitly) a proxy metric that correlates an order of functions with the compression achieved by an LZ scheme.
Suppose we are given some data to compress, e.g., a sequence of bytes representing the instructions in a binary. 
Define a \df{k-mer} to be a contiguous substring in the data of length $k$, which is a small constant. Let
$w$ be the size of the sliding window utilized by the compression algorithm; typically, $w$ is much smaller than
the length of the data. Then the compression ratio achieved by an LZ-based data compression algorithm is dictated
by the number of distinct k-mers in the data within each sliding window of size $w$. In other words, the compressed
size of the data is minimized when each k-mer is present in as few windows of size $w$ as possible.

\begin{figure}[!tb]
  \centering
  \begin{subfigure}[t]{0.49\columnwidth}
    \includegraphics[width=\columnwidth]{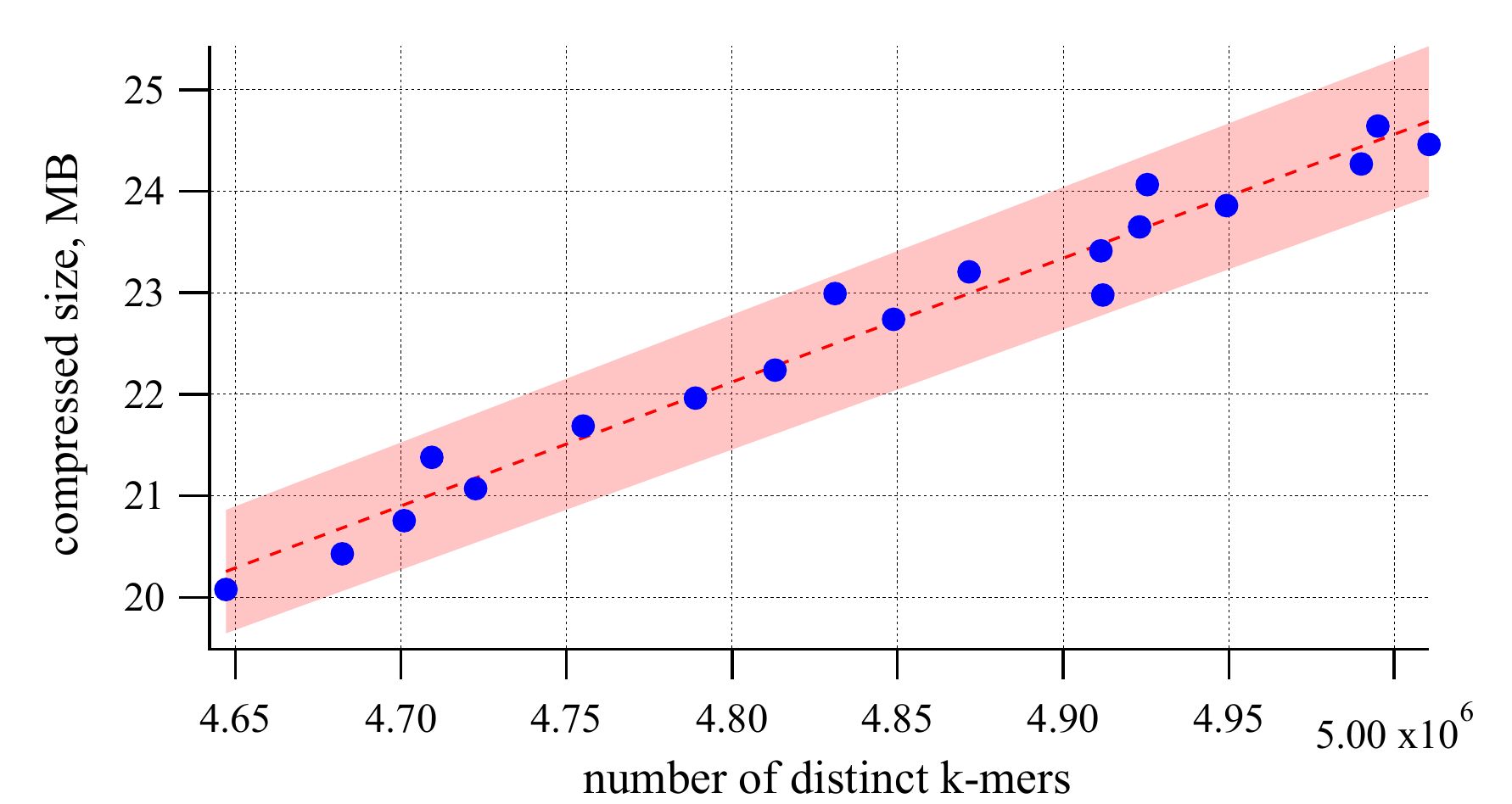}
    \caption{compressing \texttt{SocialApp} with \texttt{zstd} (level $5$)}
    \label{fig:corrA}
  \end{subfigure}
  \hfill
  \begin{subfigure}[t]{0.49\columnwidth}
    \includegraphics[width=\columnwidth]{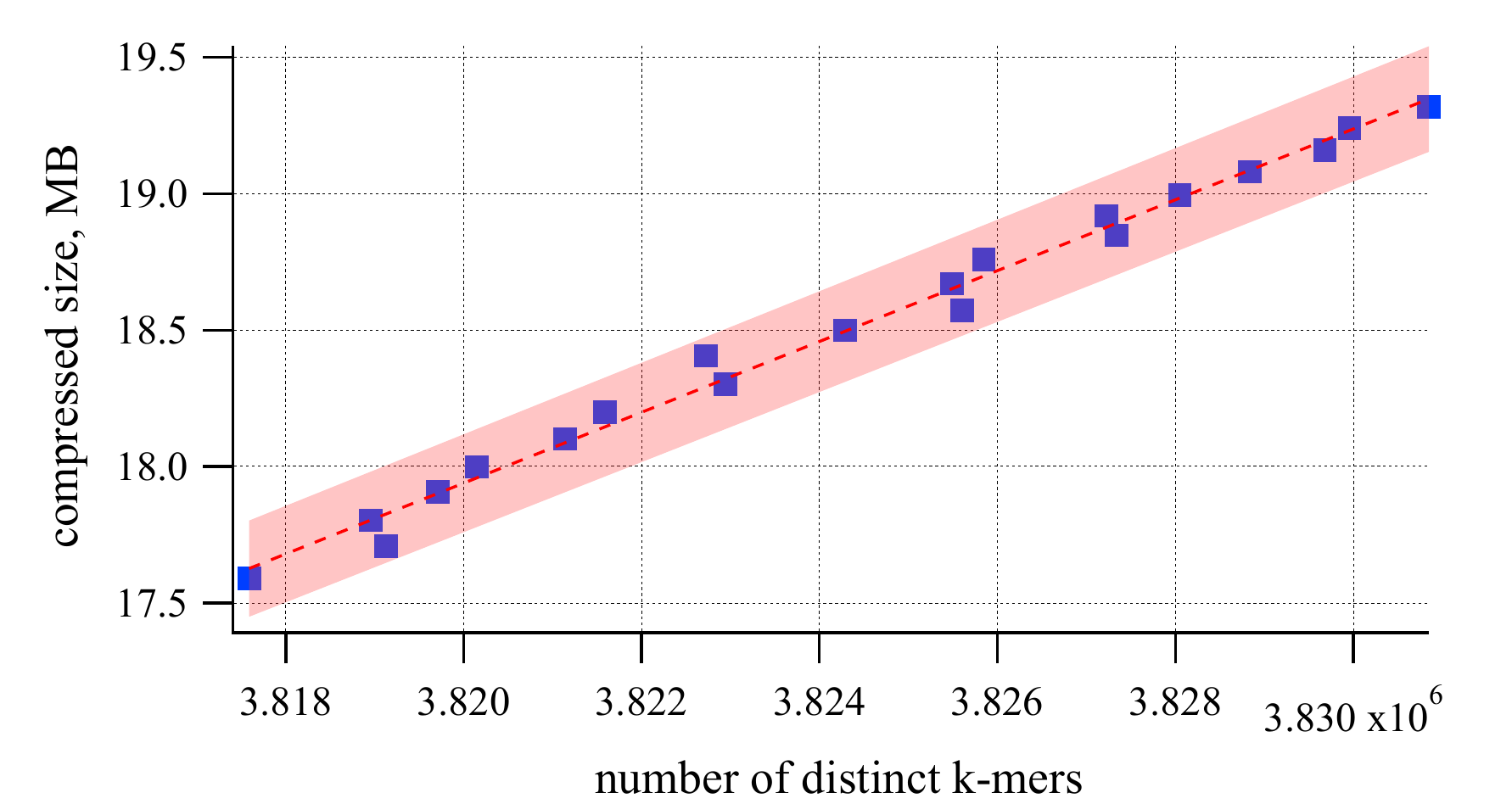}
    \caption{compressing \texttt{ChatApp} with \texttt{lz4} (level $5$)}
    \label{fig:corrB}
  \end{subfigure}
  \caption{The correlation between the number of distinct k-mers ($k=8$) in a sliding window of size $w=64$KB
    in a binary and its compressed size after applying a Lempel-Ziv-based compression algorithm}
  \label{fig:corr2}
\end{figure}

To verify the intuition, we computed and plotted the number of distinct $8$-mers within $64$KB-windows on a collection
of functions from \texttt{SocialApp} and \texttt{ChatApp}; see Figure~\ref{fig:corr2}.
To get one data point for the plot, we fixed a certain layout of functions in the binary and
extracted its \texttt{.text} section to a string, by concatenating their instructions. Then for every (contiguous) substring
of length $w$, we count the number of distinct k-mers in the substring; this number
is the proxy metric predicting the compressed size of the data. Next we apply a compression algorithm for
the entire array and measure the compressed size. To get multiple points on Figure~\ref{fig:corr2}, we
repeat the process by starting with a different function layout, which is produced from the original one
by randomly permuting some of its functions.
The results in Figure~\ref{fig:corr2} indicate a high correlation between the actual compression ratio
achieved on the data and the predicted value based on k-mers. We record a Pearson correlation 
coefficient of $\rho > 0.95$ between the two quantities. In fact, the high correlation is observed for
various values of $k$ (in our evaluation, $4 \le k \le 12$), different window sizes ($4$KB $\le k \le$ $128$KB),
and various compression tools. In particular, we experimented with 
\textsf{ZSTD} (which combines a dictionary-matching stage with a fast entropy-coding stage), 
\texttt{LZ4} (which belongs to the LZ77 family of byte-oriented compression schemes), and
\texttt{LZMA} (a variant of a dictionary compression and as a part of the \textsf{xz} tool).

\begin{figure}[!tb]
  \centering
  \begin{subfigure}[t]{0.4\columnwidth}
    \includegraphics[page=3]{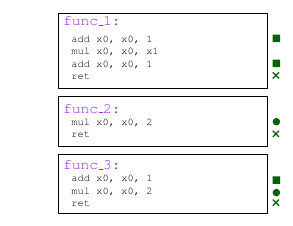}
    \caption{Functions in the binary}
    \label{fig:mcA}
  \end{subfigure}
\hfill
  \begin{subfigure}[t]{0.58\columnwidth}
    \includegraphics[page=4]{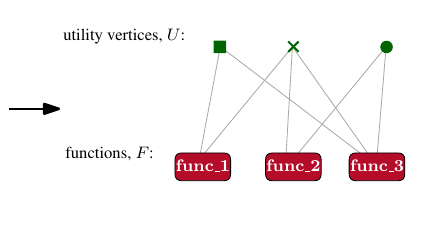}
    \caption{Bipartite graph construction}
    \label{fig:mcB}
  \end{subfigure}
  \caption{Modeling compression-aware function layout (\bpc) with a bipartite graph}
  \label{fig:model_size}
\end{figure}

Given the high predictive power of the simple proxy metric, we suggest to layout functions in a binary
so as to optimize the metric, as it can be easily extracted and computed from the data. To this end, 
we represent each function, $f \in F$, as a sequence of instructions. For every instruction in the binary, 
that occurs in at least two functions, we create a utility vertex $u \in U$.
The bipartite graph, $G=(F \cup U, E)$ contains an edge $(f, u) \in E$ if function $f$ contains instruction $u$;
refer to Figure~\ref{fig:model_size} for an illustration of the process.
The goal is to order $F$ so that instructions appear in as few windows as possible.
Equivalently, the goal is to co-locate functions that share many utility vertices, so that the
compression algorithm is able to efficiently encode the corresponding instructions.

\subsection{Start-up}
\label{sect:model:start-up}
In order to optimize the performance of the cold start, we develop the following simplified memory model. First, we assume that
when a mobile application starts, none of its code is present in the main memory. Thus, the executed code needs to be fetched from
the disk to the main memory. The disk-memory transfer happens at the granularity of memory \df{pages}, whose size is typically larger
than the size of a function. Second, we assume that when a page is fetched, it is never evicted from the memory. That is, when
a function is executed for the first time, its page should be in the memory; otherwise we get a \df{page fault}, which incurs
a start-up delay. Therefore, our goal is to find a layout of functions that results in as few page faults as possible, for
a typical start-up scenario.

In this model, the start-up performance is affected only by the first execution of a function; all subsequent executions
do not result in page faults. Hence, we record, for each function $f \in F$, the timestamp when it was
first executed, and collect the sequence of functions ordered by the timestamps.
Such sequence of functions is called the function \df{trace}. The traces list the functions participating in the
cold start and may differ from each other depending on the user or the usage scenario of the application.
Next we assume that we have a representative collection of traces, $S$.

Given an order of functions, we determine which memory page every function belongs to; to this end, we
need the sizes of the functions and assume a certain size of a page. Then for every start-up trace, $\sigma \in S$,
and an index $t \le |\sigma|$, we define $p_{\sigma}(t)$ to be the number of page faults during the execution of 
the first $t$ functions in $\sigma$. Similarly, for a set of traces $S$, we define the \df{evaluation curve} as
the average number of page faults for each $\sigma \in S$, that is, 
$p(t) := \sum_{\sigma \in S} p_{\sigma}(t) / |S|$.

\begin{figure}[!tb]
  \centering
  \begin{subfigure}[t]{0.59\columnwidth}
    \includegraphics[height=4.5cm,page=2]{pics/models}
    \caption{Bipartite graph construction with a single threshold}
    \label{fig:msA}
  \end{subfigure}
  \begin{subfigure}[t]{0.4\columnwidth}
    \includegraphics[height=4.5cm,page=1]{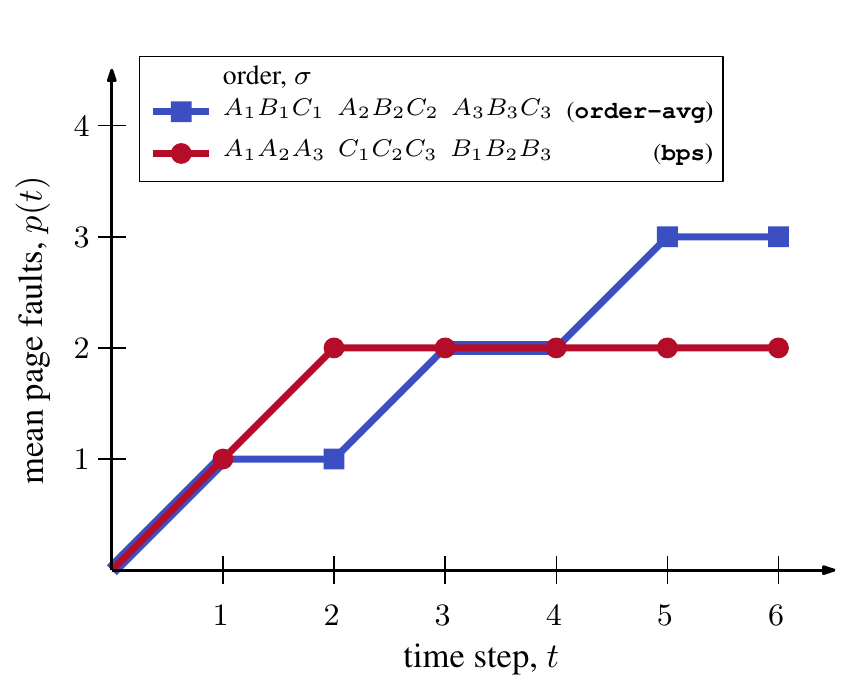}
    \caption{Evaluation curves for two orders: \bps and \alg{order-avg} suggested in~\cite{LHT22}}
    \label{fig:msB}
  \end{subfigure}
  \caption{Modeling start-up-aware function layout (\bps) with a bipartite graph}
  \label{fig:model_start}
\end{figure}

To build an intuition, consider what happens when there is a single trace $\sigma \in S$ (or equivalently, when
all traces are identical). Then the optimal layout is to use the order induced by $\sigma$, in which case
the evaluation curve looks linear in $t$. In contrast, a random permutation of functions results in
fetching most pages early in the execution and the corresponding evaluation curve looks like a step function. We 
refer the reader to Figure~\ref{fig:msB} for a concrete example of how different layouts lead to different evaluation curves. In practice, we have a diverse set of traces, and the goal is to create a function order
whose evaluation curve is as flat as possible.

We remark that although all traces have the same length, the prefix of each trace that corresponds to the startup part of that execution may differ in length due to device- and user-specific diverging execution paths. 
Therefore, instead of optimizing the value of $p(t)$ for a particular value of $t$, we try to minimize the area under the curve $p(t)$. To achieve this, we select a discrete set 
of \df{threshold} values $t_1, t_2, \ldots t_k$, and and use the bipartite graph $G = (F \cup U, E)$ with utility vertices \[U = \{(\sigma, t_i) : \sigma \in S \text{ and } 1 \leq i \leq k \},\] and edge set
\[E = \{ (f, (\sigma, t_i)) : \sigma^{-1}(f) \leq t_i \}, \]
where $\sigma^{-1}(f)$ is the index of function $f$ in $\sigma$. That way, the algorithm tries to find 
an order of $F$ in which the first $t_i$ positions of every $\sigma \in S$ occur, as much as possible, consecutively.

\section{Recursive Balanced Graph Partitioning}
\label{sect:alg}

\begin{figure}[!tb]
  \centering
  \begin{subfigure}[t]{0.48\columnwidth}
    \includegraphics[width=\columnwidth,page=2]{pics/bp2}
    \caption{The bipartite graph representation (top) and an optimization goal (bottom)} 
    \label{fig:bpA}
  \end{subfigure}
  \hfill
  \begin{subfigure}[t]{0.48\columnwidth}
    \includegraphics[width=\columnwidth,page=1]{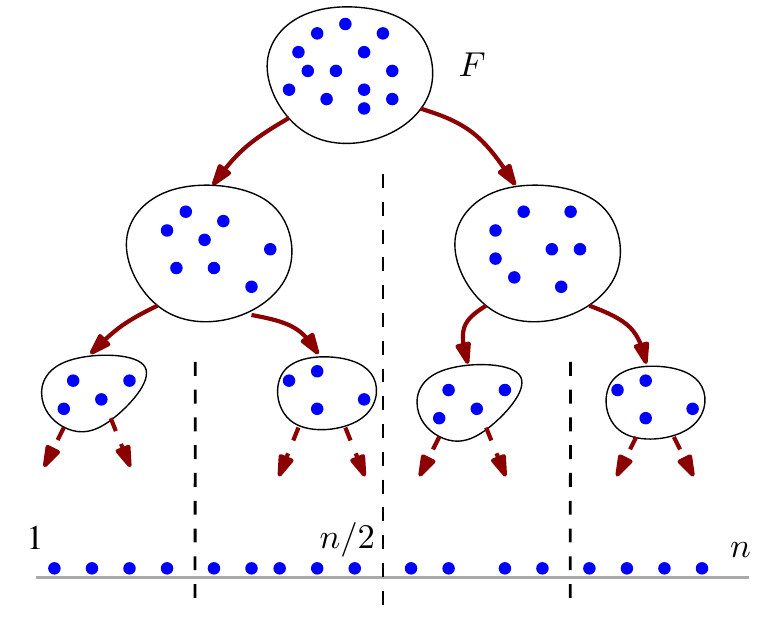}
    \caption{Recursive computation of function orders}
    \label{fig:bpB}
  \end{subfigure}
  \caption{An overview of the recursive balanced graph partitioning algorithm}
  \label{fig:bp}
\end{figure}

Our algorithm for function layout is based on the recursive balanced graph partitioning scheme. 
Recall that the input is an undirected bipartite graph 
$G=(F \cup U, E)$, where $F$ and $U$ are disjoint sets of functions and utilities, respectively, and
$E$ are the edges between them; see Figure~\ref{fig:bpA}.
The goal of the algorithm is to find a permutation of $F$ so that a certain objective is optimized.

For a high-level overview of our method, refer to Algorithm~\ref{algo:bp}.
It combines recursive graph bisection with a local search optimization on each step.
Given an input graph $G$ with $|F|=n$, we apply the bisection algorithm to obtain 
two disjoint sets of (approximately) equal cardinality, 
$F_1, F_2 \subseteq F$ with $|F_1|=\lfloor n/2 \rfloor$ and $|F_2|=\lceil n/2 \rceil$. 
We shall lay out $F_1$ on the set $\{1, \dots, \lfloor n/2 \rfloor\}$ and
lay out $F_2$ on the set $\{\lceil n/2 \rceil, \dots, n\}$. 
Thus, we have divided the problem into two problems
of half the size, and we recursively compute orders for the two subgraphs induced by vertices $F_1$
and $F_2$, adjacent utility vertices, and incident edges.
Of course, when the graph contains only one function, the order is trivial; see Figure~\ref{fig:bpB}
for an overview of the approach.

\begin{algorithm}[!t]
  \caption{Recursive Balanced Graph Partitioning}
  \label{algo:bp}
  
  \SetKwInOut{Input}{Input}
  \SetKwInOut{Output}{Output}
  \SetKwProg{Fn}{Function}{}{end}
  \SetKwFunction{ReorderBP}{ReorderBP}  
  \SetKwFunction{ComputeMoveGain}{ComputeMoveGain}  
  \Input{graph $G=(F \cup U, E)$}
  \Output{order of $F$ vertices}
  \BlankLine
  
  \Fn(){\ReorderBP}{  
    \tcc{Initial splitting of the functions into two halves.}
    \For{$f \in F$}{
      \leIf{$random(0, 1) < 0.5$}{$F_1 \leftarrow F_1 \cup \{f\}$}{$F_2 \leftarrow F_2 \cup \{f\}$}
    }
    \tcc{Refinement of the split.}
    \Repeat{converged {\bf or} iteration limit exceeded}{
      \For{$f \in F$}{$gains[f] \leftarrow \ComputeMoveGain(f)$}
      $S_1 \leftarrow $ sorted $F_1$ in descending order of $gains$\;
      $S_2 \leftarrow $ sorted $F_2$ in descending order of $gains$\;
      \For{$v \in S_1$, $u \in S_2$}{
        \If{$gains[v]+gains[u] > 0$}{
          exchange $v$ and $u$ in the sets\;}
        \lElse{break}
      }
    }
    \tcc{Recursively reorder the two parts and concatenate the results.}
    $Order_1 \leftarrow \ReorderBP(\text{Graph induced by } F_1 \cup U)$\; 
    $Order_2 \leftarrow \ReorderBP(\text{Graph induced by } F_2 \cup U)$\; 
    \Return{concatenation of $Order_1$ and $Order_2$}
  }

  \BlankLine
  \Fn(){\ComputeMoveGain{$f$}}{
    \tcc{Calculate cost improvement after moving $f$ to another part.}
    $gain = 0$\;
    \For{$(u, f) \in E$}{
      \If{$f \in F_1$}{
        $gain \leftarrow gain + cost\big(L(u), R(u)\big) - cost\big(L(u)-1, R(u)+1\big)$\;}
      \Else{
        $gain \leftarrow gain + cost\big(L(u), R(u)\big) - cost\big(L(u)+1, R(u)-1\big)$\;}       
    }
    \Return{$gain$}
  }
\end{algorithm}


Every bisection step of Algorithm~\ref{algo:bp} is a variant of the local search optimization; it is
inspired by the popular Kernighan-Lin heuristic~\cite{KL70} for the graph bisection problem. 
Initially we arbitrarily split $F$ into two sets, $F_1$ and $F_2$, and then apply a series of
iterations that exchange pairs of vertices in $F_1$ and $F_2$ trying to improve a certain \df{cost}.
To this end we compute, for every function $f \in F$, the \df{move gain}, that is, the
difference of the cost after moving $f$ from its current set to another one.
Then the vertices of $F_1$ ($F_2$) are sorted 
in the decreasing order of the gains to produce list $S_1$ ($S_2$). Finally,
we traverse the lists $S_1$ and $S_2$ in the order and exchange
the pairs of vertices, if the sum of their move gains is positive.
Note that unlike the classic graph bisection heuristic~\cite{KL70},
we do not update move gains after every swap.
The process is repeated until a convergence criterion is met (e.g., no swapped vertices)
or the maximum number of iterations is reached.
The final order of the functions is obtained by concatenating the two (recursively computed) orders for $F_1$ and $F_2$.

\paragraph*{Optimization objective}
An important aspect of the algorithm is the objective to optimize at a bisection step. Our goal is to find a layout
in which functions having many utility vertices in common are co-located in the order. We capture this goal with
the following \df{cost} of a given partition of $F$ into $F_1$ and $F_2$: 
\begin{equation}
\label{eq:cost}
cost(F_1, F_2) := \sum_{u \in U} cost\big(L(u), R(u)\big),
\end{equation}
where $L(u)$ and $R(u)$ are the numbers of functions adjacent to utility vertex $u$ in parts $F_1$ and $F_2$, respectively;
see Figure~\ref{fig:bpA}. Observe that $L(u)+R(u)$ is the degree of vertex $u$, and thus, it is independent of the split.
The objective, which we try to \emph{minimize}, is the summation of the individual contributions to the cost
over the utilities. The contribution of one utility vertex, $cost\big(L(u), R(u)\big)$, is minimized when
$L(u)=0$ or $R(u)=0$, that is, when all functions of $u$ belong to the same part; in that case, the algorithm
might be able to group the functions in the final order.
In contrast, when $L(u) \approx R(u)$, the cost takes its highest value, as the functions will likely be spread out in
the order.
Of course, it is easy to minimize the cost for one utility vertex (by placing its functions to one of the parts). However,
minimizing $cost(F_1, F_2)$ for all utilities simultaneously is a challenging task, due to the constraint on 
the sizes of $F_1$ and $F_2$.

There are multiple ways of defining the objective, $cost\big(L(u), R(u)\big)$, that satisfy the conditions above. After an 
extensive evaluation of various candidates, we identified the following objective:
\begin{equation}
\label{eq:log}  
cost\big(L(u), R(u)\big) := -L(u) \cdot \log{\big(L(u)+1\big)} - R(u) \cdot \log{\big(R(u)+1\big)},
\end{equation}
which is inspired by the so-called \df{uniform log-gap cost} utilized in the context of index compression~\cite{CKLMPR09,DKKOPS16}.
We refer to Section~\ref{sect:eval_bp} for an evaluation of alternatives and their impact on function layout.

It is straightforward to implement
$ComputeMoveGain(f)$ method from Algorithm~\ref{algo:bp} by traversing all
the edges $(u, f) \in E$ for $f \in F$ and summing up the cost differences of
moving $f$ to another set.

\paragraph*{Computational complexity}
In order to estimate the computational complexity of Algorithm~\ref{algo:bp} and predict its running time,
denote $|F| = n$ and $|E| = m$. Suppose that at each bisection step, we apply a constant
number of refinement steps (referred to as the \emph{iteration limit} in the pseudocode).
There are $\lceil \log n \rceil$ levels of recursion, and we assume that
every call of $\ReorderBP$ splits the graph into two equal-sized parts with $n/2$ vertices
and $m/2$ edges. Each call of the graph bisection consists of computing move gains and sorting
two arrays with $n$ elements. The former can be done in $\Oh(m)$ steps, while the latter takes $\Oh(n \log n)$ steps.
Therefore, the total number of steps is expressed as follows:
\[
T(n, m) = \Oh(m) + \Oh(n \log n) + 2 \cdot T(n / 2, m / 2).
\]
It is easy to verify that summing over all subproblems yields $T(n, m) = \Oh(m \log n + n \log^2 n)$, 
which is the complexity of Algorithm~\ref{algo:bp}.
We complete the discussion by noticing that it is possible to reduce the bound to $\Oh(m \log n)$
via a modified procedure for performing swaps~\cite{MPM22} but the strategy does not result
in a runtime reduction on our datasets.

The next section describes important details of our implementation.

\subsection{Algorithm Engineering}
\label{sect:eng}

While implementing Algorithm~\ref{algo:bp} in an open source compiler, we developed a few
modifications improving certain aspects of the technique. In particular, we implement the algorithm
in parallel manner and limit the depth of the recursion in order to reduce the running time.
Next we enhance the procedures for the initial split of functions into two parts and for the way
to perform swaps between the parts, which is beneficial for the quality of the identified solutions. 
Finally, we propose a sampling technique to reduce the space requirements of the algorithm.

\paragraph*{Improving the running time.}

Due to the simplicity of the algorithm, it can be implemented to run in parallel.
Notice that two subgraphs arising at the end of the bisection step are disjoint,
and thus, the two recursive calls are independent and can be processed in parallel.
To this end, we employ the fork-join computation model in which small enough
graphs are processed sequentially, while larger graphs which occur on the first few 
levels of recursion are solved in a parallel manner.

To speed up the algorithm further, we limit the depth of the recursive tree by a specified 
constant ($16$ in our implementation) and apply at most a constant number of local search iterations per 
split ($20$ in our implementation). If we reach the lowest node in the recursive tree and still 
have unordered functions, which happens when $|F| > 2^{16}$, then we use the original relative order
of the functions provided by the compiler.

Finally, we observe that the objective cost employed in the optimization requires repeated computation
of $\log(x+1)$ expressions for integer arguments. To avoid calls to evaluate floating-point logarithms, 
we create a table with pre-computed values for $0 \le x < 2^{14}$, where the upper bound is chosen
small enough to fit in the processor data cache. That way, we replaced the majority of the logarithm
evaluations with a table lookup, saving approximately $10\%$ of the total runtime.

\paragraph*{Optimizing the quality.}
One ingredient of Algorithm~\ref{algo:bp} is how the two initial sets, $F_1$ and $F_2$,
are initialized. Arguably the initialization procedure might affect the quality of 
the final vertex order, since it serves as the starting point for the local search optimization.
To initialize the bisection, we consider two alternatives. The simpler, outlined in the
pseudocode, is to randomly split $F$ into two (approximately) equal-sized sets. A more involved
strategy is to employ \df{minwise hashing}~\cite{CKLMPR09,DKKOPS16} to order the functions
by similarity and then assign the first $\lfloor n/2 \rfloor$ functions to $F_1$ and the last $\lceil n/2 \rceil$
to $F_2$. As discussed in Section~\ref{sect:eval_bp}, splitting the vertices randomly is
our preferred option due to its simplicity. 

Another interesting aspect of Algorithm~\ref{algo:bp} is the way functions exchanged between the two sets.
Recall that we pair the functions in $F_1$ with functions in $F_2$ based on the computed move gains, 
which are positive values when a function should be moved to another set or negative when a function should stay in its current set.
We observed that it is beneficial to skip some of the moves in the exchange process and keep the
function in its current set. To this end, we introduce a fixed probability ($0.1$ in our implementation) of
skipping the move for a vertex, which otherwise would have been placed to a new set.
Intuitively this adjustment prevents the optimization from becoming stuck at a local minimum that is
worse than the global one. It is also helpful for avoiding redundant swapping cycles, which might 
occur in the algorithm; refer to \cite{WS19,MPM22} for a detailed discussion of the problem
in the context of graph reordering.

\paragraph*{Reducing the space complexity.}
One potential downside to our start-up function layout algorithm is the need to collect full traces during profiling. If too many executions are profiled, then the storage requirements may be too high to be practical. A natural way to address this issue is to cap the number of traces stored, say we only store $\ell$ traces, for a fixed integer $\ell$. If the profiling process generates more than $\ell$ traces, we select a representative random sample of size $\ell$. This can be done on the fly (without knowing a priori how many traces will be generated) using reservoir sampling~\cite{Vitter85}: When the $i$th trace arrives, if $i \leq \ell$ we keep the trace, otherwise, with probability $1- \ell/ i$ we ignore the trace, and with complementary probability we pick uniformly at random one of the traces that are currently in the sample and swap it out with the new trace. The process yields a sample of $\ell$ traces chosen uniformly at random from the stream of traces.

The parameter, $\ell$, allows us to trade-off the space needed for profiling and the quality of the layout we ultimately produce (the more samples the better the layout). Figure~\ref{fig:start-up-sample} provides an empirical evaluation of this trade-off
and suggests $\ell = 300$ to be our default setting.

\section{Implementation in LLVM}
\label{sect:impl}

Both \bpc and \bps use profile data to guide function layout.
Ideally, profile data should accurately represent common real-world scenarios.
The current instrumentation in LLVM~\cite{LA04} produces an instrumented binary with large size and performance overhead due to added instrumentation instructions, added metadata sections, and changes in optimization passes.
On mobile devices, increased code size can lead to performance regressions that can change the behavior of the application.
Profiles collected from these instrumented binaries might not accurately represent our target scenarios.
To cope with this, the work of Machine IR Profile (MIP)~\cite{LHT22} focuses on small binary size and performance overhead for instrumented binaries.
This is done in part by extracting instrumentation metadata from the binary and using it to post-process the profiles offline.

MIP collects profiles that are relevant for optimizing mobile apps.
Similar to its upstream LLVM counterpart, MIP records function call counts which can be used to identify functions as either hot or cold.
MIP also collects instrumentation data that is not found in upstream LLVM.
Within each function, MIP can derive full boolean coverage data for each basic block.
MIP has an optional mode, called return address sampling, which adds probes to callees to collect 
a sample of their callsites. This can be used to construct a dynamic call graph that includes dynamically dispatched calls 
like to Objective-C's \texttt{objc\_msgSend} function.
Furthermore, MIP collects function timestamps by recording and incrementing a global timestamp for each 
function when it is called for the first time.
We sort the functions by their initial call time to construct a function trace.
To collect raw profiles at runtime, we run instrumented apps under normal usage, and dump raw 
profiles to the disk, which is uploaded to a database.
These raw profiles are later merged offline into a single optimization profile.

\subsection{Overview of the Build Pipeline}
\label{sec:impl:overview}
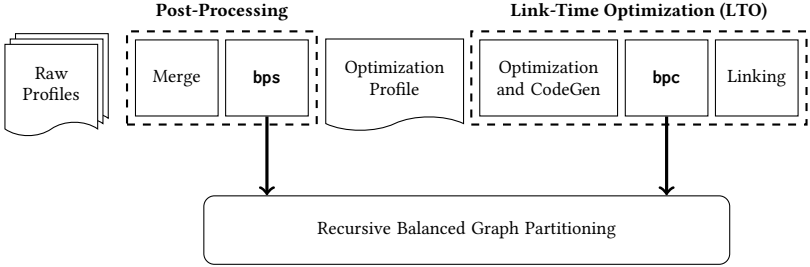
\begin{figure}[!tb]
  \centering

  \begin{tikzpicture}
    \tikzstyle{profile}=[tape,tape bend top=none,draw,font=\scriptsize, text centered,fill=white,minimum width=1cm,minimum height=1.2cm]
    \tikzstyle{compilation_step}=[draw,rectangle,minimum width=1.1cm, minimum height=1cm,text centered,font=\scriptsize]
    \tikzstyle{compilation_stage}=[thick, dashed, rectangle, draw,inner sep=3pt]
    \path (0, 0) 
    -- +(2pt, 2pt) node[profile,text width=1cm] {Raw Profiles}
    -- +(0, 0) node[profile,text width=1cm] {Raw Profiles}
    -- +(-2pt, -2pt) node[profile,text width=1cm] {Raw Profiles}
      -- ++(1.6, 0) node[compilation_step] (merge) {Merge}
    -- ++(1.2, 0) node[compilation_step] (bps) {\bps}
    -- ++(1.7, 0) node[profile, text width=1.6cm] {Optimization Profile}  
    -- ++(2.05, 0) node[compilation_step, text width=1.6cm] (codegen){Optimization and CodeGen}  
    -- ++(1.57, 0) node[compilation_step] (bpc) {\bpc}  
    -- ++(1.2, 0) node[compilation_step] (link) {Linking};
    \node [compilation_stage, fit=(merge) (bps),label=above:{\scriptsize \bf Post-Processing}] {};
    \node [compilation_stage, fit=(codegen) (link),label=above:{\scriptsize \bf Link-Time Optimization (LTO)}] {};

    \node[rounded corners, draw, rectangle, minimum height=0.9cm, minimum width=7cm, text width=4cm,text centered,font=\scriptsize] (bp) at (5.46, -2) {Recursive Balanced Graph Partitioning};

    \draw[->,very thick] (bps) edge (bps|-bp.north)
               (bpc) edge (bpc|-bp.north);

  \end{tikzpicture}
  \caption{An overview of the build pipeline with the optimized function layout
 }
  \label{fig:overview_llvm}
\end{figure}

Figure~\ref{fig:overview_llvm} shows an overview of our build pipeline.
We collect thousands of raw profile data files from various uses and periodically run offline post-processing to produce a single optimization profile.
During post-processing, \bps finds the optimized order of hot functions that were profiled, which includes start-up as well as non-start-up functions.
Our apps are built with link-time optimization~(LTO or ThinLTO).
At the end of LTO, \bpc orders cold functions for highly compressed binary size.
These two orders of functions are concatenated and passed to the linker which finalizes the function layout in the binary.
Both \bps and \bpc share the same underlying implementation.

Although we could run \bps and \bpc as one optimization pass, we have two separate passes for the following reasons:
(i) Performing \bps and \bpc at the end of LTO would require carrying large amounts of function traces through the build pipeline;
(ii) It is more convenient to simulate the expected page faults for \bps offline using function traces;
(iii) Likewise, we can evaluate \bpc, in separate, for compression without considering profile data.

\subsection{Hot Function Layout}
\label{sec:impl:bps}
As shown in Figure~\ref{fig:overview_llvm}, we first merge the raw profiles into the optimization profile 
with instrumentation metadata during post-processing.
For the block coverage and dynamic call graph data, we simply accumulate them into the optimization profile on the fly.
However, in order to run \bps, we retain function timestamps from each raw profile.
We encode the sequence of indices (that is, function trace) to the functions that participate in the cold start-up, and append them to a separate section of the optimization profile.

The \bps algorithm uses function traces with thresholds, described in Section~\ref{sect:model:start-up}, to set utility vertices, and 
produces an optimized order for start-up functions.
Once \bps is completed, the embedded function traces are no longer needed, and can be removed from the optimization profile.
Third-party library functions, and outlined functions that appear later than instrumentation in the compilation pass, might not be instrumented.
To order such functions, we first check if their call sites are profiled using block coverage data.
If that is the case, they inherit the order of their first caller.
For instance, if an uninstrumented outlined function, $f_{outlined}$, is called from the profiled functions, $f_A$ and $f_B$, and \bps orders $f_A$ followed by $f_B$,
then we insert $f_{outlined}$ after $f_A$; this results in the layout 
$f_A, f_{outlined}, f_B$.

\subsection{Cold Function Layout}
\label{sec:impl:bpc}
We run \bpc without intermediate representation (IR), after optimization and code generation are finished, as shown in Figure~\ref{fig:overview_llvm}.
Although one could run the pass utilizing IR with (\textit{Full})LTO, our approach supports ThinLTO operating on subsets of modules in parallel.
During code generation, each function publishes a set of hashes that represent its contents, which are meaningful across modules.
We use one $64$-bit \df{stable hash}~\cite{LHT22} per instruction by combining hashes of its opcode and operands. That way, every instruction is
converted to a $8$-mer, that is, a substring of length $8$.
While computing stable hashes, we omit hashes of pointers and instead utilize hashes of the contents of their targets.
Unlike outliners that need to match sequences of instructions, we neither consider the order of hashes, nor the duplicates of hashes.
We only track the set of unique stable hashes, per function, as the input to \bpc.

Since hot functions are already ordered, we filter out those functions before running \bpc.
Notice that outliners may optimistically produce many identical functions~\cite{LHT22}, which will be folded later by the linker.
To efficiently model the deduplication, \bpc groups functions that have identical sets of hashes, and runs with the set of unique functions that are cold.

As a result of \bps and \bpc, all functions are ordered.
We directly pass these orders to the linker during the LTO pipeline, and the linker enforces the layout of functions.
To avoid name conflicts for local symbols, our build pipeline uses unique naming for internal symbols.
This way, we achieve a smaller compressed app for faster downloads, while launching the app faster.

\section{Evaluation}
\label{sect:eval}

We design our experiments to answer two primary questions: 
(i)~How well does the new function layout impact real-world binaries in comparison with
alternative techniques? 
(ii)~How do various parameters of the algorithm
contribute to the solution, and what are the best parameters?
We also investigate the scalability of our algorithm.

\subsection{Experimental Setup}
\label{sec:evalsetup}
We evaluated our approach on two commercial iOS applications and one commercial Android application; refer to Table~\ref{table:data}
for basic properties of the apps. \textsf{SocialApp} is one of the largest mobile applications in the world. The app, whose total size is over $250$MB, provides a variety of usage
scenarios, which makes it an attractive target for compiler optimizations. 
\textsf{ChatApp} is a medium sized mobile app whose total size is over $50$MB.
\textsf{AndroidNative} consists of around $400$ shared natives binaries.
Unlike the two iOS binaries that are built with ThinLTO, each Android native binary is relatively small, and hence, can be compiled with (\textit{Full})LTO without significant increase in the build time.
Since there is no fully automated MIP pipeline for building \textsf{AndroidNative}, we use the app only to evaluate the compressed binary size.

%

\begingroup
\setlength{\textfloatsep}{-2pt} 
\begin{table}[!tb]
  \small
  \centering
  \caption{Basic properties of evaluated applications}
  \label{table:data}
  \begin{tabular}{lrrrrrrrr|ll}
    \toprule
    \centering 
            & \multicolumn{1}{c}{text size}
    & \multicolumn{1}{c}{binary size}
    & \multicolumn{1}{c}{total} & \multicolumn{1}{c}{hot}
    & \multicolumn{3}{c}{\underline{blocks per func.}} & \multicolumn{1}{c}{language} & \multicolumn{1}{c}{build} \\
    & (MB) & (MB) & func. &  func. & p50 & p95 & p99 & & \multicolumn{1}{c}{mode} \\
    \midrule
    \textsf{SocialApp}  &  $119$ & $259$ & $856K$  & $154K$ & $1$ & $11$ & $29$ & Obj-C/Swift & Oz+ThinLTO \\
    \textsf{ChatApp}  &  $35$ & $58$ & $202K$  & $44K$ & $3$ & $24$ & $70$ & Obj-C/C++ & Oz+ThinLTO \\
  \midrule
            \textsf{AndroidNative}  &  $38$ & $62$ & $186K$  & N/A & $3$ & $36$ & $113$ & C/C++ & Oz+LTO \\
    \bottomrule
  \end{tabular}
\end{table}
\endgroup

The experiments presented in this section were conducted on a Linux-based server with 
a dual-node 28-core 2.4 GHz Intel Xeon E5-2680 (Broadwell) having $256$GB RAM.
The algorithms are implemented on top of \texttt{release\_14} of LLVM.

\subsection{Start-up Performance}
\label{sect:eval_start}

Here we present the impact of function layout on start-up performance. 
The new algorithm, which we refer to as \bps, is compared with the following alternatives:

\begin{itemize}[leftmargin=6mm]
  \item \alg{baseline} is the original ordering functions dictated by the compiler; the function layout 
  follows the order of object files that are passed into the linker;
  
  \item \alg{random} is a result of randomly permuting the hot functions;
  
  \item \alg{order-avg} is a natural heuristic for ordering hot functions suggested in \cite{LHT22} based on the average timestamp of a function during start-up computed across all traces.
\end{itemize}

To evaluate the impact results of function layout in a production environment, we switched between using the 
current \alg{order-avg} algorithm and 
the new \bps algorithm for two different release versions, release $N$ and release $N+1$, and recorded the number 
of page faults during start-up.
Table~\ref{table:start-upprod} presents the detailed results for the number of page faults on average and $99\%$ of millions of samples published in production.
The experiment is designed this way, as
in the production environment for iOS apps, we can ship only a single binary and we cannot regress the performance by 
utilizing less effective algorithms (e.g., \alg{baseline} or \alg{random}).
We acknowledge that the improvements might come as a result of multiple optimizations, simultaneously shipped with \bps.
To account for this, we repeated the alternations three times in consecutive releases, and recorded the overall reduction 
in page faults.
On average, \bps reduced the number of major page faults by $6.9\%$ and $16.9\%$ for \textsf{SocialApp} and \textsf{ChatApp}, respectively.
The improvement translates into $4.2\%$ (average) and $2.9\%$ (p99) reductions of the cold start-up time
for \textsf{SocialApp}.

\begin{table}[!tb]
  \small
  \centering
  \caption{The number of major page faults measured for \alg{order-avg} and \bps shipped in consecutive releases.
  	The relative improvement of \bps over \alg{order-avg} is shown in parentheses.} 
  \label{table:start-upprod}
  \begin{tabular}{llll}
    \toprule
    \centering 
    & & \multicolumn{1}{c}{average} & \multicolumn{1}{c}{p99} \\
    \midrule
    \textsf{SocialApp}        & &  & \\
    \quad\alg{order-avg} & release $N$        &  $3.4K$ & $7.6K$    \\
    \quad\bps            & release $N+1$      &  $3.1K~(6.9\%)$ & $7.2K~(4.6\%)$    \\
    \textsf{ChatApp}        & &   & \\
    \quad\alg{order-avg} & release $N$        &  $1.7K$ & $10.3K$    \\
    \quad\bps & release $N+1$     &  $1.4K~(16.9\%)$ & $\:\:9.3K~(9.1\%)$   \\
    \bottomrule
  \end{tabular}
\end{table}

\begingroup
\begin{table}[!tb]
  \small
  \centering
  \caption{The relative improvements of major page faults of various function layout algorithms over \alg{baseline} measured on 
  	$iPhone12~Pro$ during the first $10$s of the cold start-up; negative values indicate regressions.}
  \label{table:start-uplocal}
  \begin{tabular}{lrr}
    \toprule
    \centering 
    & \multicolumn{1}{c}{Text} & \multicolumn{1}{c}{Binary}  \\
    \midrule
    \textsf{SocialApp}        &   & \\
    \quad\alg{random}       & $-4.7\%$ & $-3.3\%$  \\
    \quad\alg{order-avg}    & $14.1\%$ & $3.2\%$ \\
    \quad\bps           &   $19.9\%$ &  $6.8\%$  \\
    \textsf{ChatApp}        &   &  \\
    \quad\alg{random}       & $-144.4\%$ & $-65.7\%$  \\
    \quad\alg{order-avg}    & $34.9\%$ & $15.9\%$ \\
    \quad\bps           &   $36.6\%$ &  $18.3\%$  \\
    \bottomrule
  \end{tabular}
\end{table}
\endgroup

Table~\ref{table:start-uplocal} shows a similar evaluation of the start-up performance with different function layouts on a particular device, $iPhone12~Pro$, during the first $10$s of the cold start-up.
We repeat the experiment three times and calculated the mean number of page faults in (i)~the \texttt{.text} segment
and (ii)~the entire binary, which additionally includes other data segments.
Overall, \bps reduces the total major page faults in the binary by $6.8\%$ and $18.3\%$ for \textsf{SocialApp} and \textsf{ChatApp}, respectively, while \alg{order-avg} reduces them by $3.2\%$ and $15.9\%$, respectively.
As expected, \alg{random} significantly increases the page faults, negatively impacting the start-up performance.

One interesting observation from Tables~\ref{table:start-upprod}~and~\ref{table:start-uplocal} is that 
function layout has a bigger impact on the start-up performance of \textsf{ChatApp} than that of \textsf{SocialApp}.
Our explanation is that \textsf{SocialApp} consists of dozens of native binaries and four of them participate in the
start-up execution, which causes extra page faults while dynamically loading the binaries due to rebase or rebind.
In contrast, \textsf{ChatApp} consists of only a few native binaries and only one large binary is responsible for the
start-up; thus, an optimized function layout can directly impact the performance.

\begin{figure}[!tb]
  \centering
  \begin{subfigure}[t]{0.49\columnwidth}
    \includegraphics[width=\columnwidth]{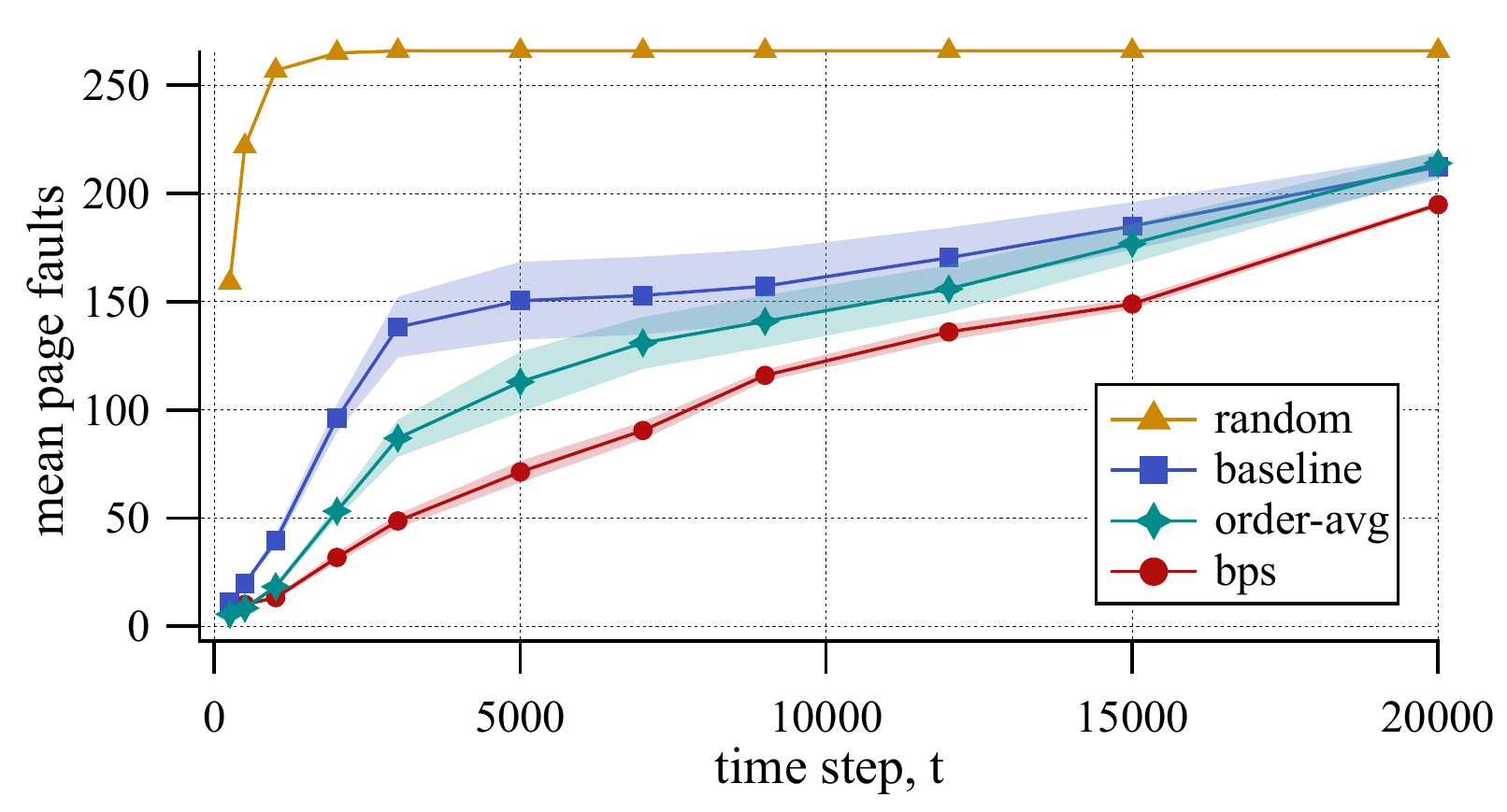}
    \caption{\label{fig:start-up-layouts}} 
  \end{subfigure}
  \hfill
  \begin{subfigure}[t]{0.49\columnwidth}
    \includegraphics[width=\columnwidth]{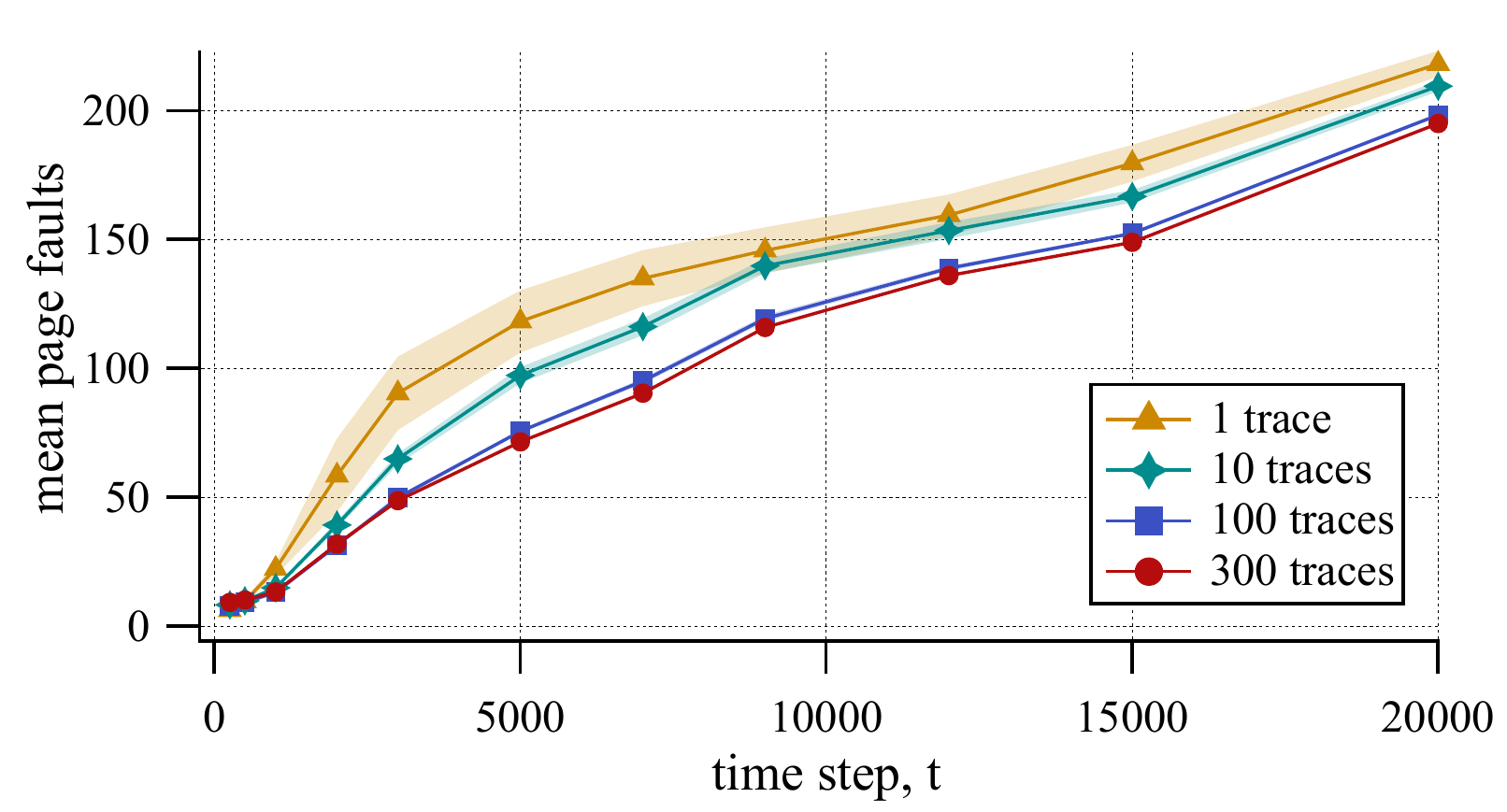}
    \caption{\label{fig:start-up-sample}}
  \end{subfigure}
  \caption{(a)~Simulated start-up performance of various function layout algorithms, and (b)~the number of sampled traces on \textsf{SocialApp}; the page size is assumed to be $16$KB with a total of $265$ memory pages}
  \label{fig:start-up}
\end{figure}

Next we investigate the start-up performance using a model developed in Section~\ref{sect:model:start-up}.
Figure~\ref{fig:start-up-layouts} illustrates the mean number of page faults simulated during start-up with up to $20$K time steps, utilizing different function layout algorithms, on \textsf{SocialApp}.
As expected, \alg{random} immediately suffers from many page faults early in the execution, as the randomized placement of functions likely spans many pages.
Interestingly, \alg{baseline} outperforms \alg{random}; this is likely due to the natural co-location of related functions
in the source code.
Then, \alg{order-avg} improves the evaluation curve over \alg{baseline}, and \bps stretches the curve even closer to the 
linear line, further minimizing the expected number of page faults.

Figure~\ref{fig:start-up-sample} shows the average number of page faults for a different number of function traces on 
\textsf{SocialApp}. Ideally, we want to build a function layout utilizing as few traces as possible.
In practice, the start-up scenarios are fairly consistent across different usages, and one can reduce the need of capturing 
all raw function traces.
We observe that running \bps with more than $300$ traces does not improve the page faults on our dataset.
As described in Section~\ref{sect:eng}, we use the value by default for \bps to reduce the space complexity of post-processing.

\subsection{Compressed Binary Size}
\label{sect:eval_size}
We now present the results of size optimizations on the selected applications. In addition to the 
\alg{baseline} and \alg{random} function layout algorithms, we compare \bpc with the following heuristic:

\begin{itemize}[leftmargin=6mm]
  \item \alg{greedy} is a greedy approach for ordering cold functions discussed in \cite{LHT22}.
  It is a procedure that iteratively builds an order by appending one function at a time. On
  each step, the most recently placed function is compared (based on the instructions)
  with not yet selected functions, and the one with the highest similarity score is appended to the
  order. To avoid an expensive $\Oh(n^2)$-computation of the scores, a number
  of pruning rules is applied to reduce the set of candidates; we refer to \cite{LHT22} for details.
\end{itemize}

Table~\ref{table:compression} summarizes the app size reduction from each function layout algorithm,
where the improvements are computed on top of \alg{baseline} (that is, the original order of 
functions generated by the compiler). The compressed size reduction is measured in three modes:
the size of the \texttt{.text} section of the binary directly impacted by our optimization, the size of the 
executables excluding resource files such as images and videos, and
the total app size in a compressed package.
We observe that \bpc reduces the size of \texttt{.text} by
$3\%$ and $1.8\%$ for \textsf{SocialApp} and \textsf{ChatApp}, respectively. Since this section
is the largest in the binary (responsible for $2/3$ of the compressed ipa size), this translates into
overall $1.9\%$ and $1.3\%$ improvements. At the same time, the impact of all the tested algorithm on the uncompressed size
of a binary is minimal (within $0.1\%$), which is mainly due to differences in code alignment.
We stress that while the absolute savings may feel insignificant,
this is a result of applying a single compiler optimization on top of the heavily tuned state-of-the-art 
techniques; the gains are comparable to
those reported by other recent works in the area~\cite{LHT22,LRN22,RPFB22,DPGPR21,LFZLS22}.

\begin{table}[!tb]
  \small
  \centering
  \caption{Compressed size improvements of various function layout algorithms over \alg{baseline}; 
    negative values indicate regressions.}
  \label{table:compression}
  \begin{tabular}{lrrrr}
    \toprule
    \centering 
    & Text & Executables & App Size & \qquad \\ 
    \midrule
    \textsf{SocialApp}        &    &   &  &   \\
    \quad\alg{random}       & $-5.3\%$ & $-4.6\%$ & $-3.7\%$ &  \\
    \quad\alg{greedy}       & $1.6\%$ & $1.3\%$ & $1.1\%$ & \\
    \quad\bpc               & $3.0\%$ & $2.3\%$ & $\boldsymbol{1.9\%}$ &  \\
    \textsf{ChatApp}        &    &   &  &   \\
    \quad\alg{random}       & $-4.9\%$ & $-4.4\%$ & $-3.8\%$ &  \\
    \quad\alg{greedy}       & $1.3\%$  & $1.0\%$ & $0.8\%$ &   \\
    \quad\bpc               & $1.8\%$  & $1.6\%$ & $\boldsymbol{1.3\%}$ &   \\
      \midrule
    \textsf{AndroidNative}        &    &   &  &   \\
    \quad\alg{random}       & $-10.0\%$ & $-8.2\%$ & $-3.2\%$ &  \\
    \quad\alg{greedy}       & $3.5\%$ & $1.9\%$ & $0.9\%$ &  \\
    \quad\bpc               & $5.2\%$ & $3.0\%$ & $\boldsymbol{1.3\%}$ &  \\
    \bottomrule
  \end{tabular}
\end{table}

An interesting observation is the behavior of \alg{random} on the dataset, which worsen the
compression ratios by around $5\%$ in comparison to the \alg{baseline} layout. Again the explanation
is that similar functions are naturally clustered in the source code. For example, 
functions within the same object file tend to have many local calls, which makes the
corresponding call instructions good candidates for a compact LZ-based encoding.
Yet \bpc is able to significantly improve the instruction locality by reordering functions
across different object files.
 
Additionally, we evaluate the compressed size reduction for \textsf{AndroidNative}.
Notice that the total app size is measured on Android package kit (apk) which includes not only native binaries, but also Android Dex bytecode.
Unlike the aforementioned two iOS apps, the \texttt{.text} size of the native binaries is only $1/4$ of the total app size.
Therefore, the overall app compressed size win is smaller than for the \texttt{.text} or the executable sections.
We observe that these compressed sizes for \textsf{AndroidNative} are more sensitive to different layouts.
This is because \textsf{AndroidNative} is a traditional C/C++ binary, where the number of blocks per function is substantially 
larger than those of the iOS apps, as illustrated in Table~\ref{table:data}.
Function call instructions encode their call targets with relative offsets whose values differ for each call-site.
Unlike the iOS apps written in Objective-C having many dynamic calls, \textsf{AndroidNative} has fewer call-sites, making it more compression-sensitive.

\subsection{Further Analysis of Balanced Partitioning}
\label{sect:eval_bp}

The new Algorithm~\ref{algo:bp} has a number of parameters that can affect its quality and performance.
In the following we discuss some of the parameters and explain our choice of their default values.

As discussed in Section~\ref{sect:alg}, the central component of the algorithm is the objective to optimize
at a bisection step, which we refer to as the \df{cost} of a partition the set of functions into 
two disjoint parts. Equation~\ref{eq:cost} provides a general form of the objective,
where the summation is taken over the utility vertices. For a given utility vertex having
$L(u)$ functions adjacent in one part and $R(u)$ functions adjacent in another part, 
$cost\big(L(u), R(u)\big)$ can be an arbitrary objective that is minimized when $L(u) = 0$ or $R(u) = 0$ and
maximized when $L(u) = R(u)$. Besides to the \df{uniform log-gap} defined by Equation~\ref{eq:log},
we consider two alternatives. The first one is \df{probabilistic fanout} defined as follows:
\[
cost\big(L(u), R(u)\big) = 1 - p^{L(u)} + 1 - p^{R(u)},
\]
where $p \in [0, 1]$ is a constant. The objective is motivated by partitioning graphs in the context
of database sharding~\cite{KKPPSAP17}, where $p$ represents the probability that a query to a database 
accesses a certain shard. The second utilized objective represents the \df{absolute difference} between $L(u)$~and~$R(u)$:
\[
cost\big(L(u), R(u)\big) = L(u) + R(u) - |L(u) - R(u)|.
\]
It is easy to verify that both objectives satisfy the requirements mentioned above.

\begin{figure}[!tb]
  \centering
  \begin{subfigure}[t]{0.48\columnwidth}
    \includegraphics[width=\columnwidth]{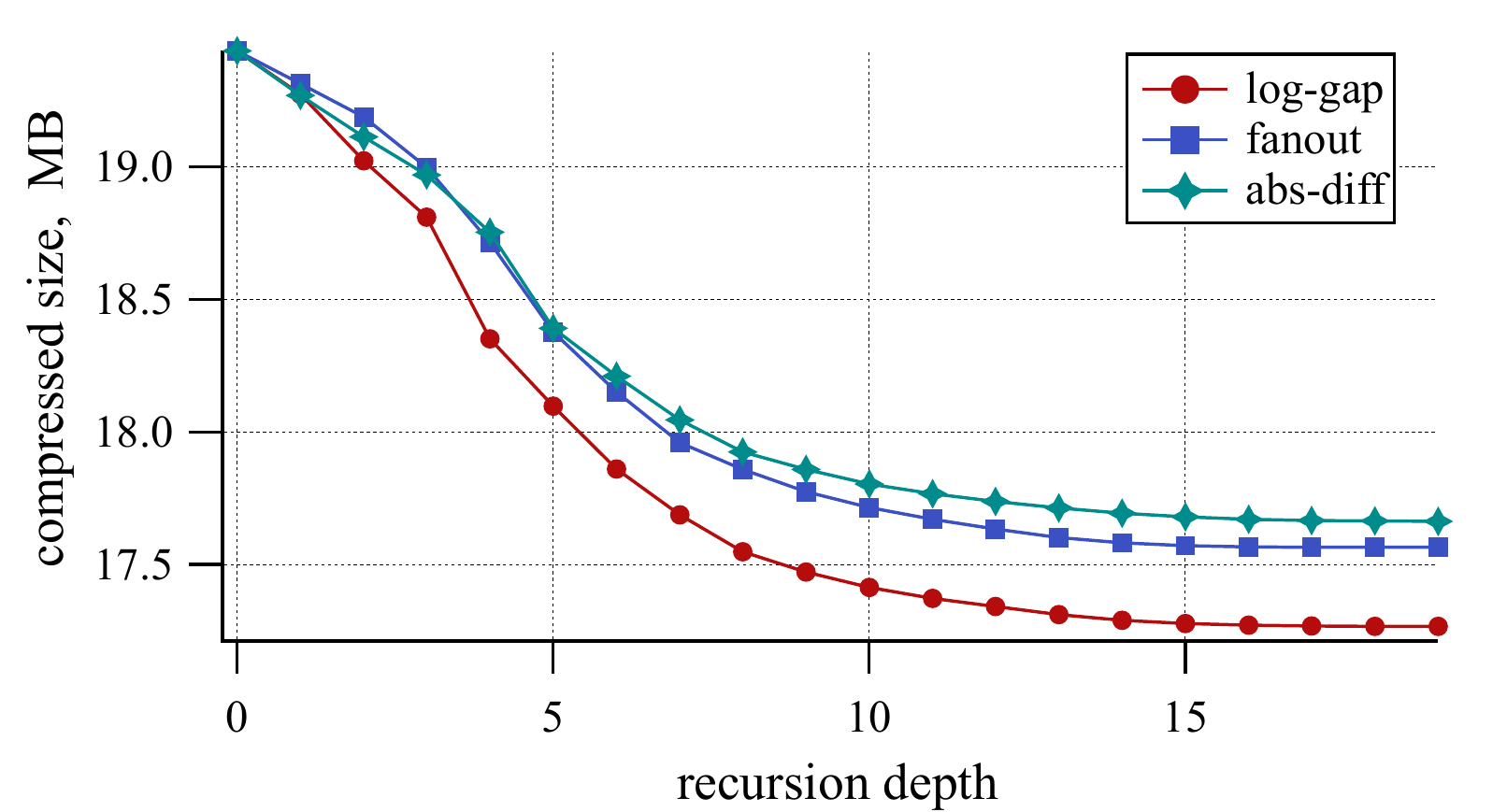}
  \end{subfigure}
  \hfill
  \begin{subfigure}[t]{0.48\columnwidth}
    \includegraphics[width=\columnwidth]{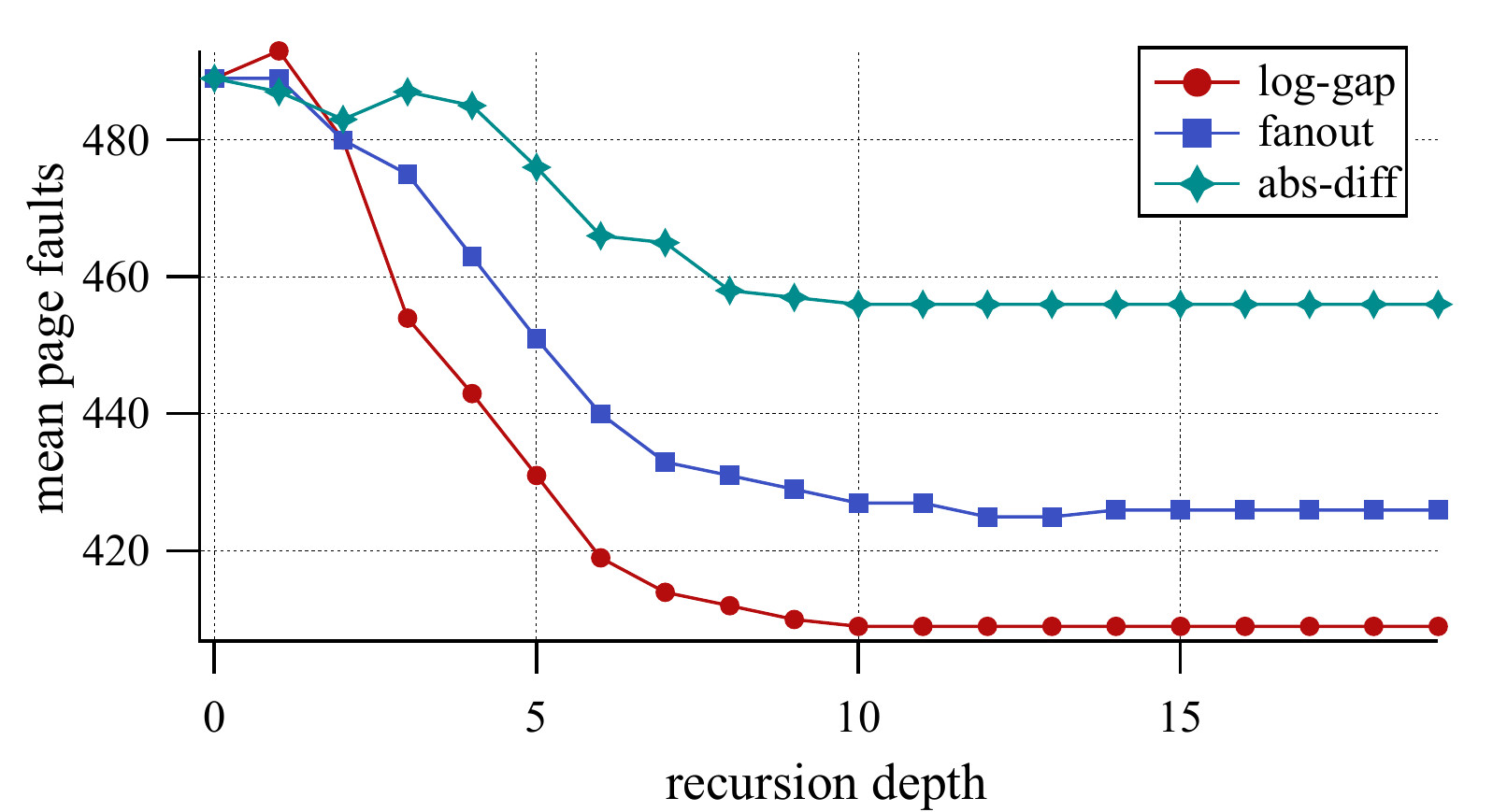}
  \end{subfigure}
  \hfill
  \begin{subfigure}[t]{0.48\columnwidth}
    \includegraphics[width=\columnwidth]{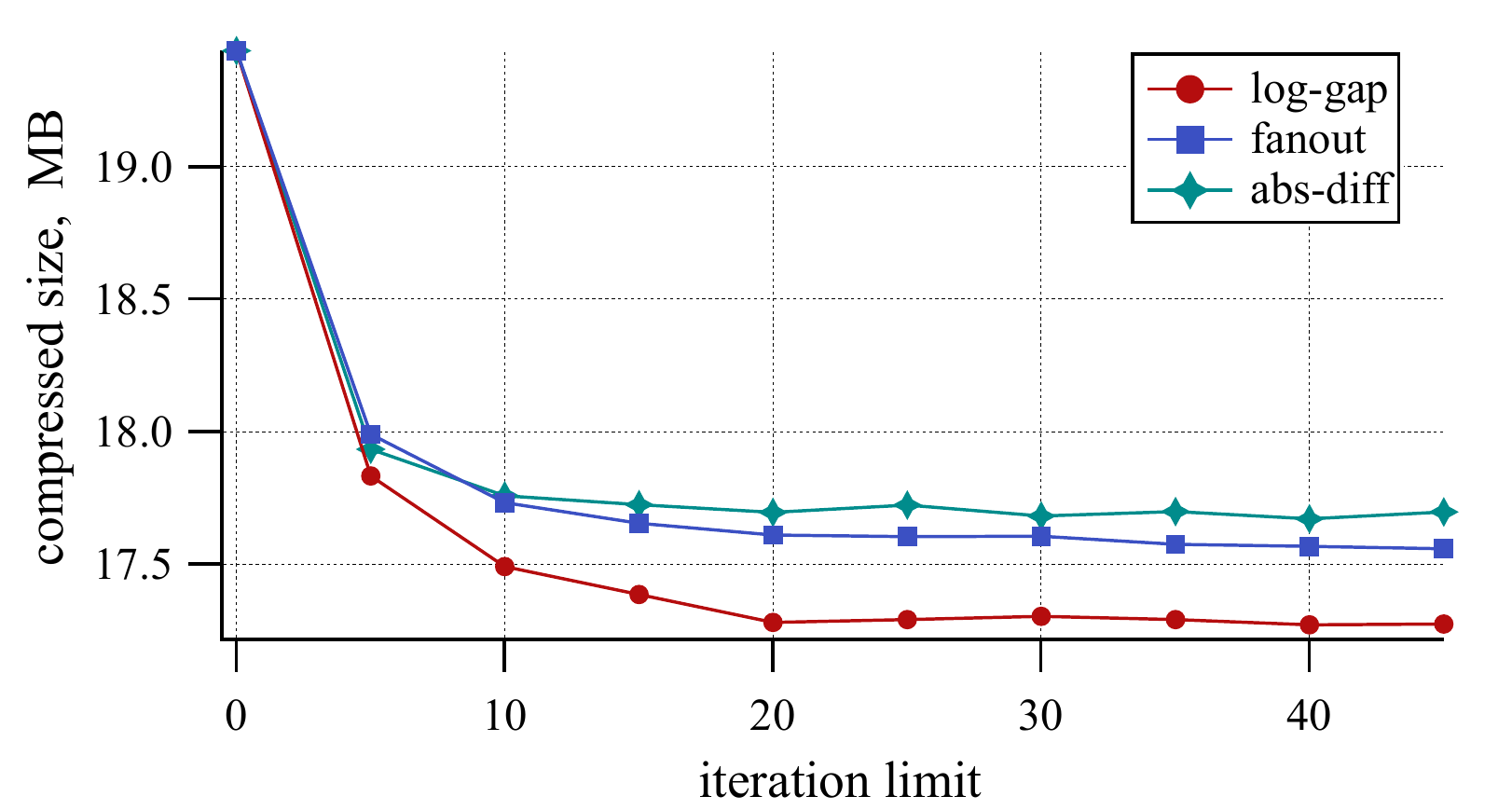}
    \caption{the application compressed size} 
  \end{subfigure}
  \hfill
  \begin{subfigure}[t]{0.48\columnwidth}
    \includegraphics[width=\columnwidth]{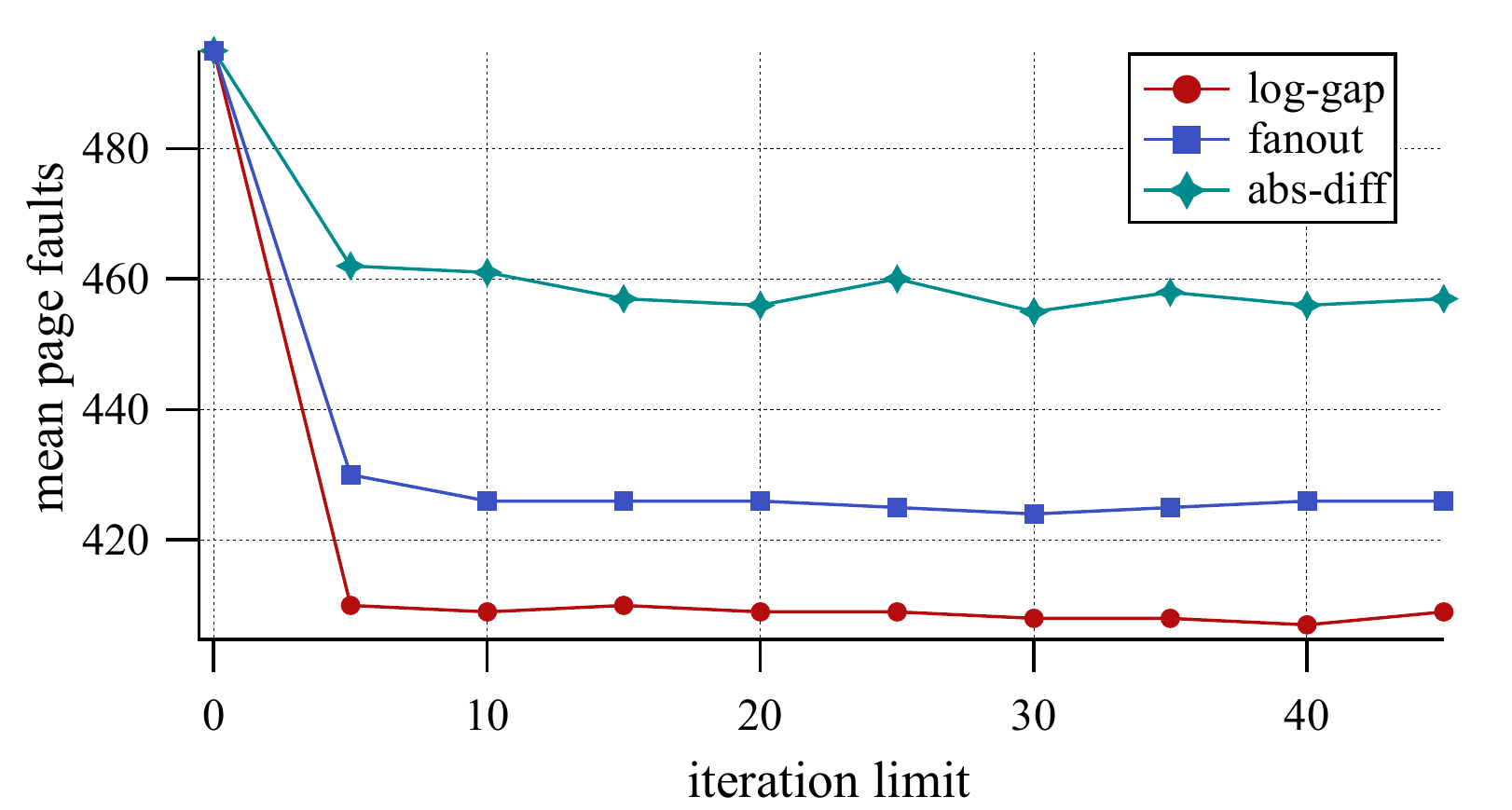}
    \caption{the application start-up time}
  \end{subfigure}
  \caption{The impact of the depth of recursion, the iteration limit, and the optimization objective
    on the quality of resulting function orders evaluated on \textsf{ChatApp}}
\vspace{-0.3cm}
  \label{fig:params}
\end{figure}

The plots on Figure~\ref{fig:params} illustrate the impact of the optimization objective on the quality
of resulting function layouts, that is, the corresponding compressed size and the (estimated) number of
page faults during start-up. For the fanout objective, we utilize $p=0.9$, as it typically results in
the best outcomes in the evaluation.
We observe that the uniform log-gap cost results in the smallest
binaries, outperforming fanout and absolute difference by $1.7\%$ and $2.3\%$, respectively. Similarly, this
objective is the best choice for start-up optimization, where it yields fewer page faults by $4.1\%$ and $11.5\%$
on average, respectively. Thus, we consider the uniform log-gap as the preferred option for the optimization.
However, function orders produced by the algorithm coupled with the other two objectives are still 
meaningfully better than alternatives investigated in Sections~\ref{sect:eval_start} and \ref{sect:eval_size}.

Next we experiment with two parameters affecting Algorithm~\ref{algo:bp}: the number of refinement iterations
and the maximum depth of the recursion. Figure~\ref{fig:params} (top) illustrates the impact of the
latter on the quality of the result evaluated on \textsf{ChatApp}. For every $0 \le d < 20$ (that is, 
when the input graph is split into $2^d$ parts), we stop the algorithm and measure the quality of
the order respecting the computed partition. It turns out that bisecting vertices is beneficial only
when $F$ contains a few tens of vertices. Therefore, we limit the depth of the recursion by
$\min(\lceil \log_2 n \rceil, 16)$ levels.
To investigate the effect of the maximum number of refinement iterations, we apply
the algorithm with various values in the range between $1$ and $45$; see Figure~\ref{fig:params} (bottom).
We observe an improvement of the quality up to iteration $20$, which serves as the default limit
in our implementation.

Finally, we explore the choice of the initial splitting strategy of $F$ into $F_1$ and $F_2$ in 
Algorithm~\ref{algo:bp}. Arguably the initialization 
procedure might affect the quality of the final order, as it provides the starting point for
the subsequent local search optimization. To verify the hypothesis,
we implemented three initialization techniques that bisect a given graph:
(i)~a random splitting as outlined in the pseudocode, (ii)~a similarity-based
minwise hashing~\cite{CKLMPR09,DKKOPS16}, and (iii)~an input-based strategy that splits
the functions based on their relative order in the compiler. In the experiments we
found no consistent winner among the three options.
Therefore, we recommend the simplest approach~(i) in the implementation.

\subsection{Build Time Analysis}
We now discuss the impact of function layout on the build time of the applications.
The time overhead by running \bpc is minimal: 
it takes less than $20$ seconds on the larger \textsf{SocialApp} and close to $1$ second on the smaller \textsf{ChatApp}.
In contrast, \alg{greedy} results in a noticeable slowdown, increasing the overall build of \textsf{SocialApp} 
by around $10-15$ minutes, which accounts for more than $10\%$ of the total build time, which is around $100$ minutes.

\begin{figure}[!tb]
  \centering
  \begin{subfigure}[t]{0.49\columnwidth}
    \includegraphics[width=\columnwidth]{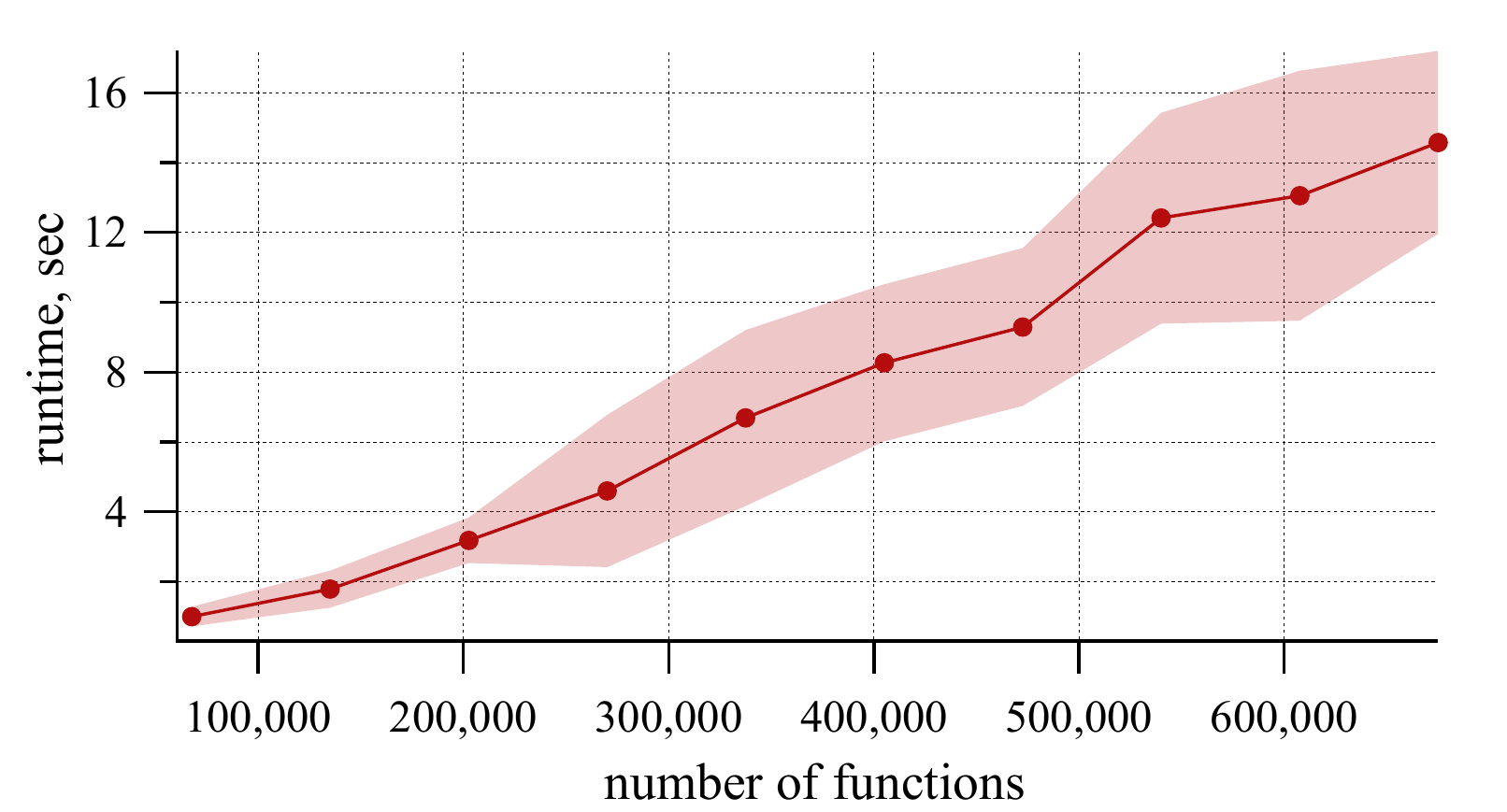}
    \caption{} 
    \label{fig:runtime_size}
  \end{subfigure}
  \hfill
  \begin{subfigure}[t]{0.49\columnwidth}
    \includegraphics[width=\columnwidth]{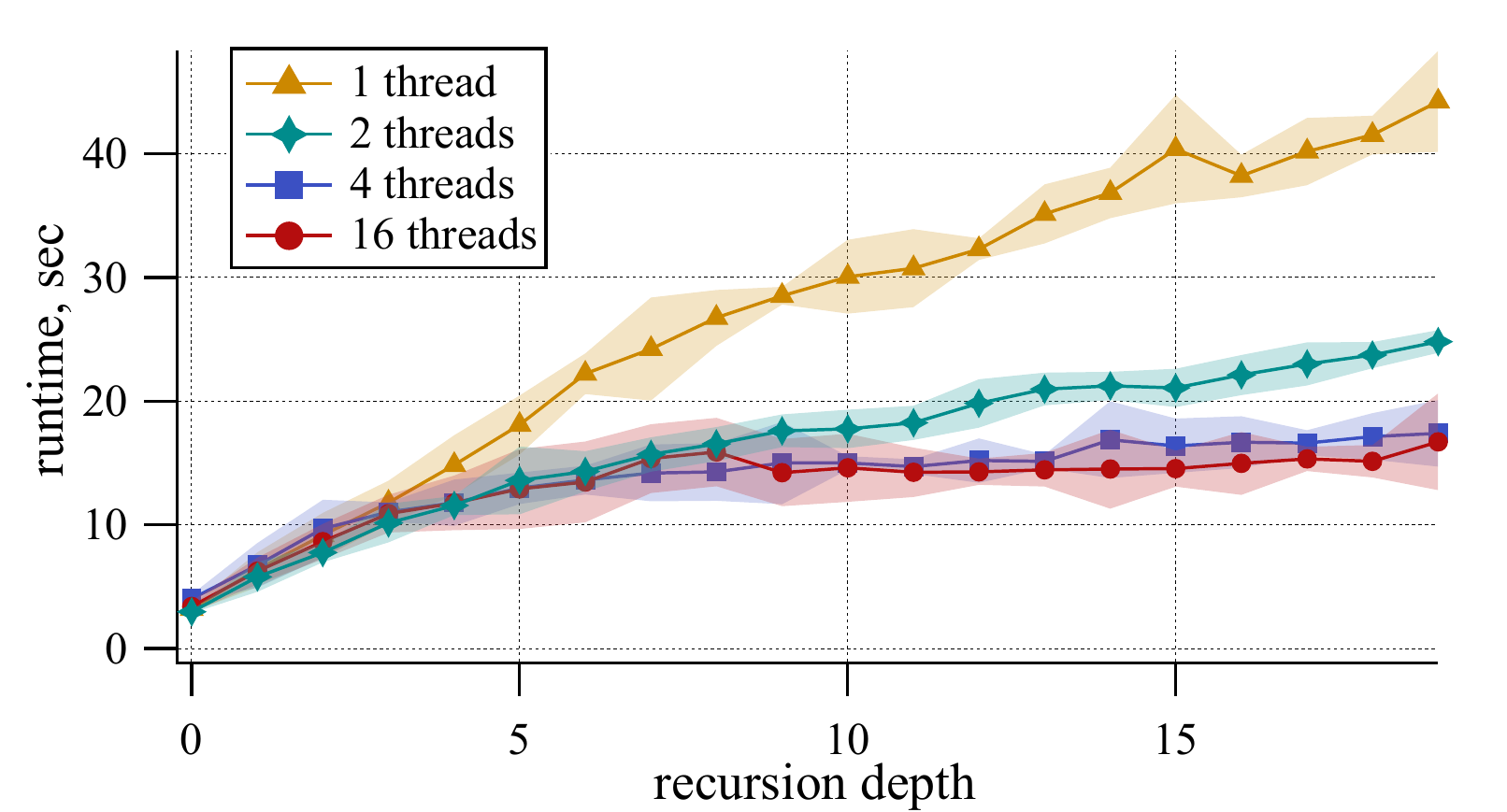}
    \caption{}
    \label{fig:runtime_threads}
  \end{subfigure}
  \caption{The impact of the number of functions and the depth of recursion on the runtime of 
    function layout evaluated on \textsf{SocialApp}}
  \label{fig:runtime}
\end{figure}

The worst-case time complexity of our implementation is upper bounded by $\Oh(m \log n + n \log^2 n)$, 
where $n$ is the total number of functions and $m$ is the number of function-utility edges.
The estimation aligns with Figure~\ref{fig:runtime_size}, which plots the dependency of the runtime 
on the number of functions in the binary.
We emphasize that the measurements are done in a multi-threaded environment in which distinct subgraphs
(arising from the recursive computation) are processed in parallel. In order to assess the speed up
of the parallelism, we limit the number of threads for the computation; see Figure~\ref{fig:runtime_threads}.
Observe that using two threads provides approximately $2$x speedup, whereas four threads yields 
a $2.5$x speedup in comparison with the single-threaded implementation. 
Increasing the number of threads beyond that does not yield measurable runtime improvements.
However it is likely that for larger instances with more recursive subgraphs, 
utilizing multiple threads can be beneficial.

\section{Related Work}
\label{sect:related}

There exists a rich literature on profile-guided compiler optimizations. 
Here we discuss previous works that are closely related to PGO in the mobile space,
code layout techniques, and our algorithmic contributions.

\paragraph{PGO}
\vspace{-0.1cm}
Most compiler optimizations for mobile applications are aimed at reducing the code size.
Such techniques include algorithms for function inlining and outlining~\cite{LRN22,DPGPR21}, 
merging similar functions~\cite{RPWCL20,RPWC21}, loop optimization~\cite{RPFB22}, unreachable
code elimination, and many others. In addition, some works describe performance
improvements for mobiles, by improving their responsiveness, memory management, and
start-up time~\cite{YCG12,LHT22}. The optimizations can be applied at the compile time or 
link time~\cite{LFZLS22}. Post-link-time PGO has been successful for large data-center applications~\cite{PANO19,Prop21}.
Our approach is complimentary to the mentioned works and can be applied in combination
with the existing optimizations.

\paragraph{Code Layout}
\vspace{-0.2cm}
The work by Pettis and Hansen~\cite{PH90} is the basis for the majority of modern code
reordering techniques for server workloads.
The goal of their basic block reordering algorithm is to create chains of blocks
that are frequently executed together in the order. Many variants of the technique
have been suggested in the literature and implemented in various tools~\cite{PANO19,OM17,NP20,Prop21,LCD19,SDAL01,MPS20}.
Alternative models have been studied in several papers~\cite{KK98,GS99,LD14}, 
where a temporal-relation graph is taken into account. Temporal affinities between code instructions can also be 
utilized for reducing conflict cache misses~\cite{HKC97}. 

Code reordering at the function-level is also initiated by Pettis and Hansen~\cite{PH90} whose 
algorithm is implemented in many compilers and binary optimization
tools~\cite{Prop21,PANO19}. This approach greedily merges chains of functions and
is designed to primarily reduce I-TLB misses. An improvement is proposed by
Ottoni and Maher~\cite{OM17}, who suggest to work with a directed call graph in order to
reduce I-cache misses. As discussed in Section~\ref{sect:intro}, the approaches are designed to 
improve the steady-state performance of server workloads and cannot be
applied for mobile apps. The very recent work of Lee, Hoag, and Tillmann~\cite{LHT22} is the
only study discussing heuristics for function layout in the mobile space; our novel algorithm significantly 
outperforms their heuristics.

\paragraph{Algorithms}
\vspace{-0.2cm}
Our model for function layout relies on the classic problem of balanced graph partitioning~\cite{AR06,GJ74,KL70}.
There exists a rich literature on graph partitioning from both theoretical and practical
points of view. We refer the reader to surveys by Bichot and Siarry~\cite{BS13} and
by Bulu{\c{c}}~et~al.~\cite{BMSSS16}. The most closely related work to our study is on graph reordering~\cite{DKKOPS16,MPM22},
which utilizes recursive graph bisection for creating ``compression-friendly'' inverted indices.
While on a high-level our algorithm shares many similarities with these works~\cite{DKKOPS16,MPM22}, the application area
and the optimization objective discussed in the paper are different.


The general problem of how to optimize memory performance has been studied from the theoretical point of view.
One classic stream of works deals with the problem of how to design cache 
eviction policies to minimize cache misses~\cite{FiatKLMSY91,SleatorT85,Young2016}. 
A more recent stream of works deals with the problem of computing a suitable data layout for 
a given cache eviction policy~\cite{PetrankR02,Lavaee16,CGOP19}. Our setting for start-up 
optimization is closest to the latter stream; however, a major difference in our setting is the 
fact that the short-time horizon of the start-up means that page evictions do not play a significant 
role in our setup. Therefore, we cannot rely on previous methods.

\section{Discussion}
\label{sect:conclude}

In this paper we designed, implemented, and evaluated the first function layout algorithm for
mobile compiler optimizations. With the careful design of the algorithm, the implementation
is fairly simple and scales to process largest instances within several seconds. We regularly apply
the optimization for large commercial mobile applications, which results in significant
start-up performance wins and app size reductions.

An important contribution of the work is a formal model for function layout optimizations. We believe 
that the model utilizing the bipartite graph with utility vertices is general enough and will be useful
in various contexts. One particularly intriguing future direction is to design a joint
optimization for hot and cold functions in the binary. In our current implementation, 
every function is either optimized for start-up or for size.
However it might be possible to relax the constraint and design an approach in which
the two objectives are unified. 
Our early experiments indicate up to $0.3\%$
size reduction when all functions are re-ordered with \bpc, possibly at the cost of a worsen start-up time. 
Unifying the optimizations is a possible future work.

From a theoretical point of view, our work is related to a computationally hard problem of
balanced graph partitioning~\cite{AR06}. While the problem is hard in theory, real-world instances
may obey certain characteristics, which may simplify the analysis of algorithms. For example, 
control-flow and call graphs arising from modern programming languages have constant \df{treewidth}, 
which is a standard notion to measure how close a graph is to a tree~\cite{CGOP19,MPS20,ADGP22}. Many 
\NP-hard optimization problems can be solved efficiently on graphs with a small treewidth, 
and therefore, exploring function layout algorithms parameterized by the treewidth is of interest.

\begin{acks}                            
We would like to thank Nikolai Tillmann for fruitful discussions of the problem,
and YongKang Zhu for helping with evaluating the approach on \textsf{AndroidNative}.
\end{acks}

\bibliographystyle{ACM-Reference-Format}
\bibliography{main-pldi}


\begin{thebibliography}{54}


\ifx \showCODEN    \undefined \def \showCODEN     #1{\unskip}     \fi
\ifx \showDOI      \undefined \def \showDOI       #1{#1}\fi
\ifx \showISBNx    \undefined \def \showISBNx     #1{\unskip}     \fi
\ifx \showISBNxiii \undefined \def \showISBNxiii  #1{\unskip}     \fi
\ifx \showISSN     \undefined \def \showISSN      #1{\unskip}     \fi
\ifx \showLCCN     \undefined \def \showLCCN      #1{\unskip}     \fi
\ifx \shownote     \undefined \def \shownote      #1{#1}          \fi
\ifx \showarticletitle \undefined \def \showarticletitle #1{#1}   \fi
\ifx \showURL      \undefined \def \showURL       {\relax}        \fi
\providecommand\bibfield[2]{#2}
\providecommand\bibinfo[2]{#2}
\providecommand\natexlab[1]{#1}
\providecommand\showeprint[2][]{arXiv:#2}

\bibitem[Ahmadi et~al\mbox{.}(2022)]%
        {ADGP22}
\bibfield{author}{\bibinfo{person}{Ali Ahmadi}, \bibinfo{person}{Majid Daliri},
  \bibinfo{person}{Amir~Kafshdar Goharshady}, {and} \bibinfo{person}{Andreas
  Pavlogiannis}.} \bibinfo{year}{2022}\natexlab{}.
\newblock \showarticletitle{Efficient approximations for cache-conscious data
  placement}. In \bibinfo{booktitle}{\emph{{PLDI} '22: 43rd {ACM} {SIGPLAN}
  International Conference on Programming Language Design and Implementation}},
  \bibfield{editor}{\bibinfo{person}{Ranjit Jhala} {and} \bibinfo{person}{Isil
  Dillig}} (Eds.). \bibinfo{publisher}{{ACM}}, \bibinfo{address}{San Diego, CA,
  USA}, \bibinfo{pages}{857--871}.
\newblock
\urldef\tempurl%
\url{https://doi.org/10.1145/3519939.3523436}
\showDOI{\tempurl}


\bibitem[Andreev and R\"{a}cke(2006)]%
        {AR06}
\bibfield{author}{\bibinfo{person}{Konstantin Andreev} {and}
  \bibinfo{person}{Harald R\"{a}cke}.} \bibinfo{year}{2006}\natexlab{}.
\newblock \showarticletitle{Balanced graph partitioning}.
\newblock \bibinfo{journal}{\emph{Theory of Computing Systems}}
  \bibinfo{volume}{39}, \bibinfo{number}{6} (\bibinfo{year}{2006}),
  \bibinfo{pages}{929--939}.
\newblock
\urldef\tempurl%
\url{https://doi.org/10.1007/s00224-006-1350-7}
\showDOI{\tempurl}


\bibitem[Bhatia(2021)]%
        {B21}
\bibfield{author}{\bibinfo{person}{Sapan Bhatia}.}
  \bibinfo{year}{2021}\natexlab{}.
\newblock \bibinfo{booktitle}{\emph{Superpack: Pushing the limits of
  compression in Facebook’s mobile apps}}.
\newblock
\urldef\tempurl%
\url{https://engineering.fb.com/2021/09/13/core-data/superpack/}
\showURL{%
\tempurl}


\bibitem[Bichot and Siarry(2013)]%
        {BS13}
\bibfield{author}{\bibinfo{person}{Charles-Edmond Bichot} {and}
  \bibinfo{person}{Patrick Siarry}.} \bibinfo{year}{2013}\natexlab{}.
\newblock \bibinfo{booktitle}{\emph{Graph Partitioning}}.
\newblock \bibinfo{publisher}{John Wiley \& Sons}.
\newblock
\urldef\tempurl%
\url{https://doi.org/10.1002/9781118601181}
\showDOI{\tempurl}


\bibitem[Bulu{\c{c}} et~al\mbox{.}(2016)]%
        {BMSSS16}
\bibfield{author}{\bibinfo{person}{Aydin Bulu{\c{c}}}, \bibinfo{person}{Henning
  Meyerhenke}, \bibinfo{person}{Ilya Safro}, \bibinfo{person}{Peter Sanders},
  {and} \bibinfo{person}{Christian Schulz}.} \bibinfo{year}{2016}\natexlab{}.
\newblock \showarticletitle{Recent Advances in Graph Partitioning}.
\newblock In \bibinfo{booktitle}{\emph{Algorithm Engineering - Selected Results
  and Surveys}}. \bibinfo{publisher}{Springer}, \bibinfo{address}{Cham},
  \bibinfo{pages}{117--158}.
\newblock
\urldef\tempurl%
\url{https://doi.org/10.1007/978-3-319-49487-6_4}
\showDOI{\tempurl}


\bibitem[Chabbi et~al\mbox{.}(2021)]%
        {CLB21}
\bibfield{author}{\bibinfo{person}{Milind Chabbi}, \bibinfo{person}{Jin Lin},
  {and} \bibinfo{person}{Raj Barik}.} \bibinfo{year}{2021}\natexlab{}.
\newblock \showarticletitle{An Experience with Code-Size Optimization for
  Production {iOS} Mobile Applications}. In
  \bibinfo{booktitle}{\emph{International Symposium on Code Generation and
  Optimization}}, \bibfield{editor}{\bibinfo{person}{Jae~W. Lee},
  \bibinfo{person}{Mary~Lou Soffa}, {and} \bibinfo{person}{Ayal Zaks}} (Eds.).
  \bibinfo{publisher}{{IEEE}}, \bibinfo{address}{Seoul, South Korea},
  \bibinfo{pages}{363--377}.
\newblock
\urldef\tempurl%
\url{https://doi.org/10.1109/CGO51591.2021.9370306}
\showDOI{\tempurl}


\bibitem[Chatterjee et~al\mbox{.}(2019)]%
        {CGOP19}
\bibfield{author}{\bibinfo{person}{Krishnendu Chatterjee},
  \bibinfo{person}{Amir~Kafshdar Goharshady}, \bibinfo{person}{Nastaran Okati},
  {and} \bibinfo{person}{Andreas Pavlogiannis}.}
  \bibinfo{year}{2019}\natexlab{}.
\newblock \showarticletitle{Efficient parameterized algorithms for data
  packing}.
\newblock \bibinfo{journal}{\emph{Proceedings of the ACM on Programming
  Languages}} \bibinfo{volume}{3}, \bibinfo{number}{POPL}
  (\bibinfo{year}{2019}), \bibinfo{pages}{1--28}.
\newblock
\urldef\tempurl%
\url{https://doi.org/10.1145/3290366}
\showDOI{\tempurl}


\bibitem[Chen et~al\mbox{.}(2016)]%
        {CML16}
\bibfield{author}{\bibinfo{person}{Dehao Chen}, \bibinfo{person}{Tipp Moseley},
  {and} \bibinfo{person}{David~Xinliang Li}.} \bibinfo{year}{2016}\natexlab{}.
\newblock \showarticletitle{{AutoFDO}: Automatic feedback-directed optimization
  for warehouse-scale applications}. In \bibinfo{booktitle}{\emph{International
  Symposium on Code Generation and Optimization}}. \bibinfo{publisher}{{ACM}},
  \bibinfo{address}{New York, NY, USA}, \bibinfo{pages}{12--23}.
\newblock
\urldef\tempurl%
\url{https://doi.org/10.1145/2854038.2854044}
\showDOI{\tempurl}


\bibitem[Chierichetti et~al\mbox{.}(2009)]%
        {CKLMPR09}
\bibfield{author}{\bibinfo{person}{Flavio Chierichetti}, \bibinfo{person}{Ravi
  Kumar}, \bibinfo{person}{Silvio Lattanzi}, \bibinfo{person}{Michael
  Mitzenmacher}, \bibinfo{person}{Alessandro Panconesi}, {and}
  \bibinfo{person}{Prabhakar Raghavan}.} \bibinfo{year}{2009}\natexlab{}.
\newblock \showarticletitle{On compressing social networks}. In
  \bibinfo{booktitle}{\emph{Knowledge Discovery and Data Mining}}.
  \bibinfo{publisher}{{ACM}}, \bibinfo{address}{Paris, France},
  \bibinfo{pages}{219--228}.
\newblock
\urldef\tempurl%
\url{https://doi.org/10.1145/1557019.1557049}
\showDOI{\tempurl}


\bibitem[Dam{\'a}sio et~al\mbox{.}(2021)]%
        {DPGPR21}
\bibfield{author}{\bibinfo{person}{Tha{\'\i}s Dam{\'a}sio},
  \bibinfo{person}{Vin{\'\i}cius Pacheco}, \bibinfo{person}{Fabr{\'\i}cio
  Goes}, \bibinfo{person}{Fernando Pereira}, {and} \bibinfo{person}{Rodrigo
  Rocha}.} \bibinfo{year}{2021}\natexlab{}.
\newblock \showarticletitle{Inlining for code size reduction}. In
  \bibinfo{booktitle}{\emph{25th Brazilian Symposium on Programming
  Languages}}. \bibinfo{publisher}{{ACM}}, \bibinfo{address}{Joinville,
  Brazil}, \bibinfo{pages}{17--24}.
\newblock
\urldef\tempurl%
\url{https://doi.org/10.1145/3475061.3475081}
\showDOI{\tempurl}


\bibitem[Developers(2022)]%
        {G22}
\bibfield{author}{\bibinfo{person}{Google Developers}.}
  \bibinfo{year}{2022}\natexlab{}.
\newblock \bibinfo{booktitle}{\emph{App Startup Time}}.
\newblock
\urldef\tempurl%
\url{https://developer.android.com/topic/performance/vitals/launch-time}
\showURL{%
\tempurl}


\bibitem[Dhulipala et~al\mbox{.}(2016)]%
        {DKKOPS16}
\bibfield{author}{\bibinfo{person}{Laxman Dhulipala}, \bibinfo{person}{Igor
  Kabiljo}, \bibinfo{person}{Brian Karrer}, \bibinfo{person}{Giuseppe
  Ottaviano}, \bibinfo{person}{Sergey Pupyrev}, {and} \bibinfo{person}{Alon
  Shalita}.} \bibinfo{year}{2016}\natexlab{}.
\newblock \showarticletitle{Compressing Graphs and Indexes with Recursive Graph
  Bisection}. In \bibinfo{booktitle}{\emph{Proceedings of the 22nd ACM SIGKDD
  International Conference on Knowledge Discovery and Data Mining}}
  \emph{(\bibinfo{series}{KDD'16})}. \bibinfo{publisher}{{ACM}},
  \bibinfo{address}{New York, NY, USA}, \bibinfo{pages}{1535--1544}.
\newblock
\urldef\tempurl%
\url{https://doi.org/10.1145/2939672.2939862}
\showDOI{\tempurl}


\bibitem[Easton and Fagin(1978)]%
        {EF78}
\bibfield{author}{\bibinfo{person}{Malcolm~C Easton} {and}
  \bibinfo{person}{Ronald Fagin}.} \bibinfo{year}{1978}\natexlab{}.
\newblock \showarticletitle{Cold-start vs. warm-start miss ratios}.
\newblock \bibinfo{journal}{\emph{Commun. ACM}} \bibinfo{volume}{21},
  \bibinfo{number}{10} (\bibinfo{year}{1978}), \bibinfo{pages}{866--872}.
\newblock


\bibitem[Ferragina and Manzini(2010)]%
        {FM10}
\bibfield{author}{\bibinfo{person}{Paolo Ferragina} {and}
  \bibinfo{person}{Giovanni Manzini}.} \bibinfo{year}{2010}\natexlab{}.
\newblock \showarticletitle{On compressing the textual web}. In
  \bibinfo{booktitle}{\emph{Proceedings of the third ACM international
  conference on Web Search and Data Mining}}. \bibinfo{publisher}{{ACM}},
  \bibinfo{address}{New York, NY, USA}, \bibinfo{pages}{391--400}.
\newblock
\urldef\tempurl%
\url{https://doi.org/10.1145/1718487.1718536}
\showDOI{\tempurl}


\bibitem[Fiat et~al\mbox{.}(1991)]%
        {FiatKLMSY91}
\bibfield{author}{\bibinfo{person}{Amos Fiat}, \bibinfo{person}{Richard~M
  Karp}, \bibinfo{person}{Michael Luby}, \bibinfo{person}{Lyle~A McGeoch},
  \bibinfo{person}{Daniel~D Sleator}, {and} \bibinfo{person}{Neal~E Young}.}
  \bibinfo{year}{1991}\natexlab{}.
\newblock \showarticletitle{Competitive paging algorithms}.
\newblock \bibinfo{journal}{\emph{Journal of Algorithms}} \bibinfo{volume}{12},
  \bibinfo{number}{4} (\bibinfo{year}{1991}), \bibinfo{pages}{685--699}.
\newblock
\urldef\tempurl%
\url{https://doi.org/10.1016/0196-6774(91)90041-V}
\showDOI{\tempurl}


\bibitem[Garey et~al\mbox{.}(1974)]%
        {GJ74}
\bibfield{author}{\bibinfo{person}{Michael~R Garey}, \bibinfo{person}{David~S
  Johnson}, {and} \bibinfo{person}{Larry Stockmeyer}.}
  \bibinfo{year}{1974}\natexlab{}.
\newblock \showarticletitle{Some simplified {NP}-complete problems}. In
  \bibinfo{booktitle}{\emph{Proceedings of the sixth annual ACM Symposium on
  Theory of Computing}}. \bibinfo{publisher}{{ACM}}, \bibinfo{address}{New
  York, NY, USA}, \bibinfo{pages}{47--63}.
\newblock
\urldef\tempurl%
\url{https://doi.org/10.1145/800119.803884}
\showDOI{\tempurl}


\bibitem[Gloy and Smith(1999)]%
        {GS99}
\bibfield{author}{\bibinfo{person}{Nikolas Gloy} {and}
  \bibinfo{person}{Michael~D Smith}.} \bibinfo{year}{1999}\natexlab{}.
\newblock \showarticletitle{Procedure placement using temporal-ordering
  information}.
\newblock \bibinfo{journal}{\emph{Transactions on Programming Languages and
  Systems}} \bibinfo{volume}{21}, \bibinfo{number}{5} (\bibinfo{year}{1999}),
  \bibinfo{pages}{977--1027}.
\newblock


\bibitem[Hashemi et~al\mbox{.}(1997)]%
        {HKC97}
\bibfield{author}{\bibinfo{person}{Amir~H Hashemi}, \bibinfo{person}{David~R
  Kaeli}, {and} \bibinfo{person}{Brad Calder}.}
  \bibinfo{year}{1997}\natexlab{}.
\newblock \showarticletitle{Efficient procedure mapping using cache line
  coloring}.
\newblock \bibinfo{journal}{\emph{SIGPLAN Notices}} \bibinfo{volume}{32},
  \bibinfo{number}{5} (\bibinfo{year}{1997}), \bibinfo{pages}{171--182}.
\newblock
\urldef\tempurl%
\url{https://doi.org/10.1145/258915.258931}
\showDOI{\tempurl}


\bibitem[He et~al\mbox{.}(2022)]%
        {HMPWY22}
\bibfield{author}{\bibinfo{person}{Wenlei He}, \bibinfo{person}{Juli{\'a}n
  Mestre}, \bibinfo{person}{Sergey Pupyrev}, \bibinfo{person}{Lei Wang}, {and}
  \bibinfo{person}{Hongtao Yu}.} \bibinfo{year}{2022}\natexlab{}.
\newblock \showarticletitle{Profile inference revisited}.
\newblock \bibinfo{journal}{\emph{Proceedings of the ACM on Programming
  Languages}} \bibinfo{volume}{6}, \bibinfo{number}{POPL}
  (\bibinfo{year}{2022}), \bibinfo{pages}{1--24}.
\newblock
\urldef\tempurl%
\url{https://doi.org/10.1145/3498714}
\showDOI{\tempurl}


\bibitem[Inc.(2022)]%
        {A22}
\bibfield{author}{\bibinfo{person}{Apple Inc.}}
  \bibinfo{year}{2022}\natexlab{}.
\newblock \bibinfo{booktitle}{\emph{Reducing Your App’s Launch Time}}.
\newblock
\urldef\tempurl%
\url{https://developer.apple.com/documentation/xcode/reducing-your-app-s-launch-time}
\showURL{%
\tempurl}


\bibitem[Inc.(2015)]%
        {F15}
\bibfield{author}{\bibinfo{person}{Facebook Inc.}}
  \bibinfo{year}{2015}\natexlab{}.
\newblock \bibinfo{booktitle}{\emph{Optimizing Facebook for {iOS} Start Time}}.
\newblock
\urldef\tempurl%
\url{https://engineering.fb.com/2015/11/20/ios/optimizing-facebook-for-ios-start-time}
\showURL{%
\tempurl}


\bibitem[Inc(2021)]%
        {Redex}
\bibfield{author}{\bibinfo{person}{Facebook Inc}.}
  \bibinfo{year}{2021}\natexlab{}.
\newblock \bibinfo{booktitle}{\emph{Redex: {A} bytecode optimizer for {A}ndroid
  apps}}.
\newblock
\urldef\tempurl%
\url{https://fbredex.com}
\showURL{%
\tempurl}


\bibitem[Kabiljo et~al\mbox{.}(2017)]%
        {KKPPSAP17}
\bibfield{author}{\bibinfo{person}{Igor Kabiljo}, \bibinfo{person}{Brian
  Karrer}, \bibinfo{person}{Mayank Pundir}, \bibinfo{person}{Sergey Pupyrev},
  \bibinfo{person}{Alon Shalita}, \bibinfo{person}{Yaroslav Akhremtsev}, {and}
  \bibinfo{person}{Alessandro Presta}.} \bibinfo{year}{2017}\natexlab{}.
\newblock \showarticletitle{Social Hash Partitioner: {A} Scalable Distributed
  Hypergraph Partitioner}.
\newblock \bibinfo{journal}{\emph{Proc. {VLDB} Endow.}} \bibinfo{volume}{10},
  \bibinfo{number}{11} (\bibinfo{year}{2017}), \bibinfo{pages}{1418--1429}.
\newblock
\urldef\tempurl%
\url{https://doi.org/10.14778/3137628.3137650}
\showDOI{\tempurl}


\bibitem[Kalamationos and Kaeli(1998)]%
        {KK98}
\bibfield{author}{\bibinfo{person}{J Kalamationos} {and}
  \bibinfo{person}{David~R Kaeli}.} \bibinfo{year}{1998}\natexlab{}.
\newblock \showarticletitle{Temporal-based procedure reordering for improved
  instruction cache performance}. In \bibinfo{booktitle}{\emph{High-Performance
  Computer Architecture}}. \bibinfo{publisher}{{IEEE} Computer Society},
  \bibinfo{address}{Las Vegas, Nevada, USA}, \bibinfo{pages}{244--253}.
\newblock
\urldef\tempurl%
\url{https://doi.org/10.1109/HPCA.1998.650563}
\showDOI{\tempurl}


\bibitem[Kernighan and Lin(1970)]%
        {KL70}
\bibfield{author}{\bibinfo{person}{Brian~W Kernighan} {and}
  \bibinfo{person}{Shen Lin}.} \bibinfo{year}{1970}\natexlab{}.
\newblock \showarticletitle{An efficient heuristic procedure for partitioning
  graphs}.
\newblock \bibinfo{journal}{\emph{Bell System Technical Journal}}
  \bibinfo{volume}{49}, \bibinfo{number}{2} (\bibinfo{year}{1970}),
  \bibinfo{pages}{291--307}.
\newblock


\bibitem[Lattner and Adve(2004)]%
        {LA04}
\bibfield{author}{\bibinfo{person}{Chris Lattner} {and} \bibinfo{person}{Vikram
  Adve}.} \bibinfo{year}{2004}\natexlab{}.
\newblock \showarticletitle{{LLVM}: A compilation framework for lifelong
  program analysis \& transformation}. In
  \bibinfo{booktitle}{\emph{International Symposium on Code Generation and
  Optimization}}. \bibinfo{publisher}{{IEEE} Computer Society},
  \bibinfo{address}{San Jose, CA, {USA}}, \bibinfo{pages}{75}.
\newblock
\urldef\tempurl%
\url{https://doi.org/10.1109/CGO.2004.1281665}
\showDOI{\tempurl}


\bibitem[Lavaee(2016)]%
        {Lavaee16}
\bibfield{author}{\bibinfo{person}{Rahman Lavaee}.}
  \bibinfo{year}{2016}\natexlab{}.
\newblock \showarticletitle{The hardness of data packing}. In
  \bibinfo{booktitle}{\emph{Proceedings of the 43rd {SIGPLAN-SIGACT} Symposium
  on Principles of Programming Languages}}. \bibinfo{publisher}{{ACM}},
  \bibinfo{address}{St. Petersburg, FL, USA}, \bibinfo{pages}{232--242}.
\newblock
\urldef\tempurl%
\url{https://doi.org/10.1145/2837614.2837669}
\showDOI{\tempurl}


\bibitem[Lavaee et~al\mbox{.}(2019)]%
        {LCD19}
\bibfield{author}{\bibinfo{person}{Rahman Lavaee}, \bibinfo{person}{John
  Criswell}, {and} \bibinfo{person}{Chen Ding}.}
  \bibinfo{year}{2019}\natexlab{}.
\newblock \showarticletitle{Codestitcher: inter-procedural basic block layout
  optimization}. In \bibinfo{booktitle}{\emph{Proceedings of the 28th
  International Conference on Compiler Construction}},
  \bibfield{editor}{\bibinfo{person}{Jos{\'{e}}~Nelson Amaral} {and}
  \bibinfo{person}{Milind Kulkarni}} (Eds.). \bibinfo{publisher}{{ACM}},
  \bibinfo{address}{Washington, DC, USA}, \bibinfo{pages}{65--75}.
\newblock
\urldef\tempurl%
\url{https://doi.org/10.1145/3302516.3307358}
\showDOI{\tempurl}


\bibitem[Lavaee and Ding(2014)]%
        {LD14}
\bibfield{author}{\bibinfo{person}{Rahman Lavaee} {and} \bibinfo{person}{Chen
  Ding}.} \bibinfo{year}{2014}\natexlab{}.
\newblock \bibinfo{booktitle}{\emph{{ABC Optimizer}: Affinity Based Code Layout
  Optimization}}.
\newblock \bibinfo{type}{{T}echnical {R}eport}.
  \bibinfo{institution}{University of Rochester}.
\newblock


\bibitem[Lee et~al\mbox{.}(2022a)]%
        {LHT22}
\bibfield{author}{\bibinfo{person}{Kyungwoo Lee}, \bibinfo{person}{Ellis Hoag},
  {and} \bibinfo{person}{Nikolai Tillmann}.} \bibinfo{year}{2022}\natexlab{a}.
\newblock \showarticletitle{Efficient profile-guided size optimization for
  native mobile applications}. In \bibinfo{booktitle}{\emph{{ACM} {SIGPLAN}
  International Conference on Compiler Construction}}.
  \bibinfo{publisher}{{ACM}}, \bibinfo{address}{Seoul, South Korea},
  \bibinfo{pages}{243--253}.
\newblock
\urldef\tempurl%
\url{https://doi.org/10.1145/3497776.3517764}
\showDOI{\tempurl}


\bibitem[Lee et~al\mbox{.}(2022b)]%
        {LRN22}
\bibfield{author}{\bibinfo{person}{Kyungwoo Lee}, \bibinfo{person}{Manman Ren},
  {and} \bibinfo{person}{Shane Nay}.} \bibinfo{year}{2022}\natexlab{b}.
\newblock \showarticletitle{Scalable size inliner for mobile applications
  (WIP)}. In \bibinfo{booktitle}{\emph{Proceedings of the 23rd ACM
  SIGPLAN/SIGBED International Conference on Languages, Compilers, and Tools
  for Embedded Systems}}. \bibinfo{publisher}{{ACM}}, \bibinfo{address}{San
  Diego, CA, USA}, \bibinfo{pages}{116--120}.
\newblock
\urldef\tempurl%
\url{https://doi.org/10.1145/3519941.3535074}
\showDOI{\tempurl}


\bibitem[Lin et~al\mbox{.}(2014)]%
        {LLDSW14}
\bibfield{author}{\bibinfo{person}{Xing Lin}, \bibinfo{person}{Guanlin Lu},
  \bibinfo{person}{Fred Douglis}, \bibinfo{person}{Philip Shilane}, {and}
  \bibinfo{person}{Grant Wallace}.} \bibinfo{year}{2014}\natexlab{}.
\newblock \showarticletitle{Migratory compression: {C}oarse-grained data
  reordering to improve compressibility}. In \bibinfo{booktitle}{\emph{USENIX
  Conference on File and Storage Technologies (FAST)}}.
  \bibinfo{publisher}{{USENIX}}, \bibinfo{address}{Santa Clara, CA, USA},
  \bibinfo{pages}{257--271}.
\newblock


\bibitem[Liu et~al\mbox{.}(2022)]%
        {LFZLS22}
\bibfield{author}{\bibinfo{person}{Gai Liu}, \bibinfo{person}{Umar Farooq},
  \bibinfo{person}{Chengyan Zhao}, \bibinfo{person}{Xia Liu}, {and}
  \bibinfo{person}{Nian Sun}.} \bibinfo{year}{2022}\natexlab{}.
\newblock \showarticletitle{Linker Code Size Optimization for Native Mobile
  Applications}.
\newblock \bibinfo{journal}{\emph{arXiv preprint arXiv:2210.07311}}
  \bibinfo{volume}{abs/2210.07311} (\bibinfo{year}{2022}).
\newblock
\urldef\tempurl%
\url{https://doi.org/10.48550/arXiv.2210.07311}
\showDOI{\tempurl}


\bibitem[Mackenzie et~al\mbox{.}(2022)]%
        {MPM22}
\bibfield{author}{\bibinfo{person}{Joel Mackenzie}, \bibinfo{person}{Matthias
  Petri}, {and} \bibinfo{person}{Alistair Moffat}.}
  \bibinfo{year}{2022}\natexlab{}.
\newblock \showarticletitle{Tradeoff Options for Bipartite Graph Partitioning}.
\newblock \bibinfo{journal}{\emph{IEEE Transactions on Knowledge and Data
  Engineering}} (\bibinfo{year}{2022}), \bibinfo{pages}{1--15}.
\newblock
\urldef\tempurl%
\url{https://doi.org/10.1109/TKDE.2022.3208902}
\showDOI{\tempurl}


\bibitem[Mansour(2020)]%
        {M20}
\bibfield{author}{\bibinfo{person}{Nezar Mansour}.}
  \bibinfo{year}{2020}\natexlab{}.
\newblock \bibinfo{booktitle}{\emph{Understanding Cold, Hot, and Warm App
  Launch Time}}.
\newblock
\urldef\tempurl%
\url{https://blog.instabug.com/understanding-cold-hot-and-warm-app-launch-time/}
\showURL{%
\tempurl}


\bibitem[Mestre et~al\mbox{.}(2021)]%
        {MPS20}
\bibfield{author}{\bibinfo{person}{Juli\'{a}n Mestre}, \bibinfo{person}{Sergey
  Pupyrev}, {and} \bibinfo{person}{Seeun~William Umboh}.}
  \bibinfo{year}{2021}\natexlab{}.
\newblock \showarticletitle{On the Extended {TSP} Problem}. In
  \bibinfo{booktitle}{\emph{32nd International Symposium on Algorithms and
  Computation}} \emph{(\bibinfo{series}{LIPIcs}, Vol.~\bibinfo{volume}{212})},
  \bibfield{editor}{\bibinfo{person}{Hee{-}Kap Ahn} {and}
  \bibinfo{person}{Kunihiko Sadakane}} (Eds.). \bibinfo{publisher}{Schloss
  Dagstuhl - Leibniz-Zentrum f{\"{u}}r Informatik}, \bibinfo{address}{Fukuoka,
  Japan}, \bibinfo{pages}{42:1--42:14}.
\newblock
\urldef\tempurl%
\url{https://doi.org/10.4230/LIPIcs.ISAAC.2021.42}
\showDOI{\tempurl}


\bibitem[Newell and Pupyrev(2020)]%
        {NP20}
\bibfield{author}{\bibinfo{person}{Andy Newell} {and} \bibinfo{person}{Sergey
  Pupyrev}.} \bibinfo{year}{2020}\natexlab{}.
\newblock \showarticletitle{Improved Basic Block Reordering}.
\newblock \bibinfo{journal}{\emph{{IEEE} Transactions in Computers}}
  \bibinfo{volume}{69}, \bibinfo{number}{12} (\bibinfo{year}{2020}),
  \bibinfo{pages}{1784--1794}.
\newblock
\urldef\tempurl%
\url{https://doi.org/10.1109/TC.2020.2982888}
\showDOI{\tempurl}


\bibitem[Ottoni and Maher(2017)]%
        {OM17}
\bibfield{author}{\bibinfo{person}{Guilherme Ottoni} {and}
  \bibinfo{person}{Bertrand Maher}.} \bibinfo{year}{2017}\natexlab{}.
\newblock \showarticletitle{Optimizing Function Placement for Large-scale
  Data-center Applications}. In \bibinfo{booktitle}{\emph{International
  Symposium on Code Generation and Optimization}}. \bibinfo{publisher}{IEEE
  Press}, \bibinfo{address}{Austin, USA}, \bibinfo{pages}{233--244}.
\newblock
\urldef\tempurl%
\url{https://doi.org/10.1109/CGO.2017.7863743}
\showDOI{\tempurl}


\bibitem[Panchenko et~al\mbox{.}(2019)]%
        {PANO19}
\bibfield{author}{\bibinfo{person}{Maksim Panchenko}, \bibinfo{person}{Rafael
  Auler}, \bibinfo{person}{Bill Nell}, {and} \bibinfo{person}{Guilherme
  Ottoni}.} \bibinfo{year}{2019}\natexlab{}.
\newblock \showarticletitle{{BOLT}: a practical binary optimizer for data
  centers and beyond}. In \bibinfo{booktitle}{\emph{International Symposium on
  Code Generation and Optimization}}. \bibinfo{publisher}{{IEEE}},
  \bibinfo{address}{Washington, DC, USA}, \bibinfo{pages}{2--14}.
\newblock
\urldef\tempurl%
\url{https://doi.org/10.1109/CGO.2019.8661201}
\showDOI{\tempurl}


\bibitem[Petrank and Rawitz(2005)]%
        {PetrankR02}
\bibfield{author}{\bibinfo{person}{Erez Petrank} {and} \bibinfo{person}{Dror
  Rawitz}.} \bibinfo{year}{2005}\natexlab{}.
\newblock \showarticletitle{The hardness of cache conscious data placement}.
\newblock \bibinfo{journal}{\emph{Nordic Journal of Computing}}
  \bibinfo{volume}{12}, \bibinfo{number}{3} (\bibinfo{year}{2005}),
  \bibinfo{pages}{275--307}.
\newblock


\bibitem[Pettis and Hansen(1990)]%
        {PH90}
\bibfield{author}{\bibinfo{person}{Karl Pettis} {and} \bibinfo{person}{Robert~C
  Hansen}.} \bibinfo{year}{1990}\natexlab{}.
\newblock \showarticletitle{Profile guided code positioning}.
\newblock \bibinfo{journal}{\emph{SIGPLAN Notices}} \bibinfo{volume}{25},
  \bibinfo{number}{6} (\bibinfo{year}{1990}), \bibinfo{pages}{16--27}.
\newblock
\urldef\tempurl%
\url{https://doi.org/10.1145/989393.989433}
\showDOI{\tempurl}


\bibitem[Propeller(2020)]%
        {Prop21}
Propeller \bibinfo{year}{2020}\natexlab{}.
\newblock \bibinfo{title}{Propeller: Profile Guided Optimizing Large Scale
  {LLVM}-based Relinker}.
\newblock
  \bibinfo{howpublished}{\url{https://github.com/google/llvmpropeller}}.
\newblock


\bibitem[Raskhodnikova et~al\mbox{.}(2013)]%
        {RRRS13}
\bibfield{author}{\bibinfo{person}{Sofya Raskhodnikova}, \bibinfo{person}{Dana
  Ron}, \bibinfo{person}{Ronitt Rubinfeld}, {and} \bibinfo{person}{Adam
  Smith}.} \bibinfo{year}{2013}\natexlab{}.
\newblock \showarticletitle{Sublinear algorithms for approximating string
  compressibility}.
\newblock \bibinfo{journal}{\emph{Algorithmica}} \bibinfo{volume}{65},
  \bibinfo{number}{3} (\bibinfo{year}{2013}), \bibinfo{pages}{685--709}.
\newblock
\urldef\tempurl%
\url{https://doi.org/10.1007/s00453-012-9618-6}
\showDOI{\tempurl}


\bibitem[Reinhardt(2016)]%
        {R16}
\bibfield{author}{\bibinfo{person}{Peter Reinhardt}.}
  \bibinfo{year}{2016}\natexlab{}.
\newblock \bibinfo{booktitle}{\emph{Effect of Mobile App Size on Downloads}}.
\newblock
\urldef\tempurl%
\url{https://segment.com/blog/mobile-app-size-effect-on-downloads/}
\showURL{%
\tempurl}


\bibitem[Rocha et~al\mbox{.}(2022)]%
        {RPFB22}
\bibfield{author}{\bibinfo{person}{Rodrigo~CO Rocha}, \bibinfo{person}{Pavlos
  Petoumenos}, \bibinfo{person}{Bj{\"o}rn Franke}, \bibinfo{person}{Pramod
  Bhatotia}, {and} \bibinfo{person}{Michael O’Boyle}.}
  \bibinfo{year}{2022}\natexlab{}.
\newblock \showarticletitle{Loop rolling for code size reduction}. In
  \bibinfo{booktitle}{\emph{IEEE/ACM International Symposium on Code Generation
  and Optimization (CGO)}}, \bibfield{editor}{\bibinfo{person}{Jae~W. Lee},
  \bibinfo{person}{Sebastian Hack}, {and} \bibinfo{person}{Tatiana Shpeisman}}
  (Eds.). \bibinfo{publisher}{{IEEE}}, \bibinfo{address}{Seoul, Republic of
  Korea}, \bibinfo{pages}{217--229}.
\newblock
\urldef\tempurl%
\url{https://doi.org/10.1109/CGO53902.2022.9741256}
\showDOI{\tempurl}


\bibitem[Rocha et~al\mbox{.}(2021)]%
        {RPWC21}
\bibfield{author}{\bibinfo{person}{Rodrigo~CO Rocha}, \bibinfo{person}{Pavlos
  Petoumenos}, \bibinfo{person}{Zheng Wang}, \bibinfo{person}{Murray Cole},
  \bibinfo{person}{Kim Hazelwood}, {and} \bibinfo{person}{Hugh Leather}.}
  \bibinfo{year}{2021}\natexlab{}.
\newblock \showarticletitle{HyFM: Function merging for free}. In
  \bibinfo{booktitle}{\emph{Proceedings of the 22nd ACM SIGPLAN/SIGBED
  International Conference on Languages, Compilers, and Tools for Embedded
  Systems}}, \bibfield{editor}{\bibinfo{person}{J{\"{o}}rg Henkel} {and}
  \bibinfo{person}{Xu~Liu}} (Eds.). \bibinfo{publisher}{{ACM}},
  \bibinfo{address}{Virtual Event, Canada}, \bibinfo{pages}{110--121}.
\newblock
\urldef\tempurl%
\url{https://doi.org/10.1145/3461648.3463852}
\showDOI{\tempurl}


\bibitem[Rocha et~al\mbox{.}(2020)]%
        {RPWCL20}
\bibfield{author}{\bibinfo{person}{Rodrigo~CO Rocha}, \bibinfo{person}{Pavlos
  Petoumenos}, \bibinfo{person}{Zheng Wang}, \bibinfo{person}{Murray Cole},
  {and} \bibinfo{person}{Hugh Leather}.} \bibinfo{year}{2020}\natexlab{}.
\newblock \showarticletitle{Effective function merging in the {SSA} form}. In
  \bibinfo{booktitle}{\emph{Proceedings of the 41st ACM SIGPLAN Conference on
  Programming Language Design and Implementation}},
  \bibfield{editor}{\bibinfo{person}{Alastair~F. Donaldson} {and}
  \bibinfo{person}{Emina Torlak}} (Eds.). \bibinfo{publisher}{{ACM}},
  \bibinfo{address}{London, UK}, \bibinfo{pages}{854--868}.
\newblock
\urldef\tempurl%
\url{https://doi.org/10.1145/3385412.3386030}
\showDOI{\tempurl}


\bibitem[Schwarz et~al\mbox{.}(2001)]%
        {SDAL01}
\bibfield{author}{\bibinfo{person}{Benjamin Schwarz}, \bibinfo{person}{Saumya
  Debray}, \bibinfo{person}{Gregory Andrews}, {and} \bibinfo{person}{Matthew
  Legendre}.} \bibinfo{year}{2001}\natexlab{}.
\newblock \showarticletitle{{PLTO}: A link-time optimizer for the {I}ntel
  {IA}-32 architecture}. In \bibinfo{booktitle}{\emph{Workshop on Binary
  Rewriting}}. \bibinfo{pages}{1--7}.
\newblock


\bibitem[Sleator and Tarjan(1985)]%
        {SleatorT85}
\bibfield{author}{\bibinfo{person}{Daniel~D Sleator} {and}
  \bibinfo{person}{Robert~E Tarjan}.} \bibinfo{year}{1985}\natexlab{}.
\newblock \showarticletitle{Amortized efficiency of list update and paging
  rules}.
\newblock \bibinfo{journal}{\emph{Commun. ACM}} \bibinfo{volume}{28},
  \bibinfo{number}{2} (\bibinfo{year}{1985}), \bibinfo{pages}{202--208}.
\newblock


\bibitem[Vitter(1985)]%
        {Vitter85}
\bibfield{author}{\bibinfo{person}{Jeffrey~Scott Vitter}.}
  \bibinfo{year}{1985}\natexlab{}.
\newblock \showarticletitle{Random Sampling with a Reservoir}.
\newblock \bibinfo{journal}{\emph{{ACM} Trans. Math. Softw.}}
  \bibinfo{volume}{11}, \bibinfo{number}{1} (\bibinfo{year}{1985}),
  \bibinfo{pages}{37--57}.
\newblock
\urldef\tempurl%
\url{https://doi.org/10.1145/3147.3165}
\showDOI{\tempurl}


\bibitem[Wang and Suel(2019)]%
        {WS19}
\bibfield{author}{\bibinfo{person}{Qi Wang} {and} \bibinfo{person}{Torsten
  Suel}.} \bibinfo{year}{2019}\natexlab{}.
\newblock \showarticletitle{Document reordering for faster intersection}.
\newblock \bibinfo{journal}{\emph{Proceedings of the VLDB Endowment}}
  \bibinfo{volume}{12}, \bibinfo{number}{5} (\bibinfo{year}{2019}),
  \bibinfo{pages}{475--487}.
\newblock


\bibitem[Yan et~al\mbox{.}(2012)]%
        {YCG12}
\bibfield{author}{\bibinfo{person}{Tingxin Yan}, \bibinfo{person}{David Chu},
  \bibinfo{person}{Deepak Ganesan}, \bibinfo{person}{Aman Kansal}, {and}
  \bibinfo{person}{Jie Liu}.} \bibinfo{year}{2012}\natexlab{}.
\newblock \showarticletitle{Fast app launching for mobile devices using
  predictive user context}. In \bibinfo{booktitle}{\emph{Proceedings of the
  10th international conference on Mobile systems, applications, and
  services}}. \bibinfo{publisher}{{ACM}}, \bibinfo{address}{Ambleside, United
  Kingdom}, \bibinfo{pages}{113--126}.
\newblock
\urldef\tempurl%
\url{https://doi.org/10.1145/2307636.2307648}
\showDOI{\tempurl}


\bibitem[Young(2016)]%
        {Young2016}
\bibfield{author}{\bibinfo{person}{Neal~E. Young}.}
  \bibinfo{year}{2016}\natexlab{}.
\newblock \showarticletitle{Online Paging and Caching}.
\newblock In \bibinfo{booktitle}{\emph{Encyclopedia of Algorithms}}.
  \bibinfo{publisher}{Springer New York}, \bibinfo{address}{New York, NY},
  \bibinfo{pages}{1457--1461}.
\newblock
\urldef\tempurl%
\url{https://doi.org/10.1007/978-1-4939-2864-4_267}
\showDOI{\tempurl}


\bibitem[Ziv and Lempel(1977)]%
        {LZ77}
\bibfield{author}{\bibinfo{person}{Jacob Ziv} {and} \bibinfo{person}{Abraham
  Lempel}.} \bibinfo{year}{1977}\natexlab{}.
\newblock \showarticletitle{A universal algorithm for sequential data
  compression}.
\newblock \bibinfo{journal}{\emph{IEEE Transactions on information theory}}
  \bibinfo{volume}{23}, \bibinfo{number}{3} (\bibinfo{year}{1977}),
  \bibinfo{pages}{337--343}.
\newblock


\end{thebibliography}
  
\end{document}